\documentclass{emulateapj}

\newcommand{\vc}[1]{\mbox{\boldmath{$#1$}}}

\DeclareMathSymbol{\varOmega}{\mathord}{letters}{"0A}
\DeclareMathSymbol{\varSigma}{\mathord}{letters}{"06}
\DeclareMathSymbol{\varPsi}{\mathord}{letters}{"09}

\newcommand{\Eq}[1]{equation (\ref{#1})}

\newcommand{\Fig}[1]{Fig.~\ref{#1}}
\newcommand{\Figs}[2]{Figs.~\ref{#1} and \ref{#2}}
 
\newcommand{\Tab}[1]{Table \ref{#1}}

\newcommand{\ts}{\tau_{\rm f}}
\newcommand{\ep}{\epsilon}

\shorttitle{Turbulence Driven by the Streaming Instability}
\shortauthors{Johansen \& Youdin}

\slugcomment{Accepted for publication in ApJ}

\begin{document}

\title{Protoplanetary Disk Turbulence Driven by the Streaming Instability:\\
Non-Linear Saturation and Particle Concentration}

\author{A. Johansen\altaffilmark{1}}
\affil{Max-Planck-Institut f\"ur Astronomie, 69117 Heidelberg, Germany}
\email{johansen@mpia.de}

\and

\author{A. Youdin\altaffilmark{2}}
\affil{Princeton University Observatory, Princeton, NJ 08544}
\email{youd@cita.utoronto.ca}

\altaffiltext{1}{Visiting Astronomer, Department of Astrophysics, American
Museum of Natural History}
\altaffiltext{2}{Current address: Canadian
Institute for Theoretical Astrophysics, University of Toronto, 60
Saint George Street, Toronto, ON M5S 3H8, Canada}

\begin{abstract}
  We present simulations of the non-linear evolution of streaming instabilities
  in protoplanetary disks. The two components of the disk, gas treated with
  grid hydrodynamics and solids treated as superparticles, are mutually coupled
  by drag forces. We find that the initially laminar equilibrium flow
  spontaneously develops into turbulence in our unstratified local model.
  Marginally coupled solids (that couple to the gas on a Keplerian time-scale)
  trigger an upward cascade to large particle clumps with peak overdensities
  above 100. The clumps evolve dynamically by losing material downstream to the
  radial drift flow while receiving recycled material from upstream. Smaller,
  more tightly coupled solids produce weaker turbulence with more transient
  overdensities on smaller length scales. The net inward radial drift is
  decreased for marginally coupled particles, whereas the tightly coupled
  particles migrate faster in the saturated turbulent state. The turbulent
  diffusion of solid particles, measured by their random walk, depends strongly
  on their stopping time and on the solids-to-gas ratio of the background
  state, but diffusion is generally modest, particularly for tightly coupled
  solids. Angular momentum transport is too weak and of the wrong sign to
  influence stellar accretion. Self-gravity and collisions will be needed to
  determine the relevance of particle overdensities for planetesimal formation.
\end{abstract}


\section{Introduction}

This paper extends our preparatory numerical investigations of the streaming
instability \citep[hereafter YJ]{yj07} into the non-linear regime. The linear
streaming instability was first described by \citet[hereafter YG]{yg05} who
found that the radial and azimuthal drift of solids through gas in a
protoplanetary disk triggers growing oscillations that concentrate particles.
YJ details the numerical techniques used to study the evolution of solids and
gas in a local patch of a protoplanetary disk and demonstrates that our code
successfully reproduces the linear growth rates derived by YG. This paper
describes the non-linear evolution of the streaming instability to a fully
turbulent state and studies the consequences for particle concentration and
transport.

The starward drift of solids, caused by the sub-Keplerian headwind encountered
by the particles, is not just a trigger for streaming instabilities, but also a
source of theoretical difficulties. Growing planetesimals by coagulation faces
severe time-scale constraints due to the loss off solids (ultimately to the
star or sublimation in the inner disk). The restriction is most acute for $10$
cm ``rocks" through $1$ m ``boulders" with drift times of only a few hundred
orbits \citep{Weidenschilling1977} in most of the planet-forming region of
standard minimum mass nebula models \citep{stu77b,hay81}. The drift of mm-sized
solids in a few times $10^5$ years at 30 AU is at best marginally consistent
with the observed mm-excess from the outer parts T-Tauri disks with ages of a
few Myr (\citealt{Wilner+etal2000,Rodmann+etal2006}; see also
\citealt{Brauer+etal2007} for possible ways to maintain such a population of
``pebbles"). This mismatch between theory and observations may indicate that
simple drift time estimates are missing important dynamical effects.

Several mechanisms could impede the radial influx of solids. The increased
inertia of solids in a dense midplane sublayer decreases drift speeds as the
local gas mass fraction squared \citep{Nakagawa+etal1986,yc04}. Since such high
densities may trigger rapid gravitational collapse of solids \citep{ys02},
sedimentation alone is not a satisfactory explanation of the long lifetimes of
pebbles in the disks, even if the turbulence is weak enough to allow the
formation of an extremely thin mid-plane layer. Turbulent diffusion in
accretion disks will maintain a small fraction of particles in the outer disk
\citep{sv96, tak02}, but this scenario requires a particle reservoir that
exceeds by far the observed amount of mm-sized solids and thus implies disk
masses that are orders of magnitude larger than minimum mass models. Giant
anticyclonic vortices \citep{BargeSommeria1995, delaFuenteMarcosBarge2001}
stall migration by trapping marginally coupled solids. However the formation
and stability of vortices in disks is not clear
\citep{Goodman+etal1987,kb03,jab04,bm05,FromangNelson2005}. Any local pressure
maximum -- not only vortices -- can trap boulders
\citep{kl01,HaghighipourBoss2003}, e.g.\ spiral arms of massive
self-gravitating disks \citep{Rice+etal2004} or even transient pressure
enhancements in magnetorotational turbulence \citep{JohansenKlahrHenning2006}.
The present work will show that streaming turbulence modestly slows the average
radial drift of marginally coupled solids. An ultimate solution of the drift
problem may require rapid (faster than drift) planetesimal formation (by
gravitational collapse and/or coagulation) \emph{and} fragmentation to maintain
observed populations of small solids \citep{dd05}.

The dynamical particle trapping mechanisms mentioned above increase particle
densities, with an efficiency that depends on (often uncertain) structure
lifetimes. Local particle overdensities can seed gravitational collapse of
solids and affect the rates of (and balance between) coagulation and
collisional fragmentation.  Optically thick clumps could even influence
radiative transfer if the disk itself is optically thin, and thereby alter
observational estimates of disk mass and particle size (see \citealp{Draine06}
for a general discussion of the role of optical depth, but not clumping
\emph{per se}). Radial drift inherently augments the surface density of solids
in the inner disk as particles pile up from larger orbital radii, as long as
particles are smaller than the gas mean free path so that Epstein drag applies
\citep{ys02,yc04}. In simulations and experiments of forced Kolmogorov
turbulence, particles concentrate in low vorticity regions at the viscous
dissipation scale \citep{fke94,Padoan+etal2006}. Efficient collection requires
small particles that couple to the rapid turnover time at the dissipation
scale. \citet{cuz01} apply this passive turbulent concentration to the
size-sorting of chondrules (abundant, partially-molten, mm-sized inclusions
found in primitive meteorites) in the inner solar nebula.

\cite{jhk06} discovered active turbulent concentration (active meaning that the
drag feedback on gas was included) of larger particles (from cm-sized pebbles
to m-sized boulders) in simulations of Kelvin-Helmholz midplane turbulence.
Dense clumps of solids plow through the gas at near the Keplerian speed,
scooping up more isolated particles that move with the sub-Keplerian headwind.
Since this particle concentration mechanism relies on two-way drag forces, it
was (and still is) considered a non-linear manifestation of streaming
instabilities. The current work further explores active concentration,
isolating the role of drag feedback by ignoring vertical stratification. We
consider different geometries, both axisymmetric in the radial-vertical plane
and fully 3-D, than \citet{jhk06}, who considered the azimuthal-vertical plane,
and we use the higher order interpolation scheme for drag forces described in
YJ. We will thus show that the ``pure" streaming instability also produces
strongly non-linear particle overdensities.

Turbulent diffusion controls the extent to which particles sediment in the
mid-plane \citep{Dubrulle+etal1995} and whether (or how fast) self-gravity can
collect solids into rings and bound clumps \citep{y05a,y05b}. Diffusion is the
most fundamental parameter governing whether planetesimals can form by
gravitational instability (as originally proposed by \citealp{saf69, gw73}),
because the oft-mentioned critical density for gravitational collapse is
irrelevant when drag forces are included to transfer angular momentum from the
solids to the gas \citep{war76,war00}. Passive diffusion of particles in
magnetorotational turbulence has been found to be quite strong
\citep[][]{JohansenKlahr2005,Turner+etal2006,FromangPapaloizou2006,Carballido+etal2006},
although the Schmidt number (the ratio between the turbulent viscosity of the
gas and the particle diffusion) increases in the presence of a net vertical
magnetic field \citep{Carballido+etal2005,JohansenKlahrMee2006}. In the present
work we measure the active particle diffusion in streaming turbulence, by
considering the random walk of the particles away from a reference point, and
find it to be relatively weak, especially for smaller particles. 

The paper is built up as follows. In \S\ref{c:numerics} we briefly reiterate
the physical model of the protoplanetary disk and our numerical method for
solving the dynamical equations of gas and solids. In \S\ref{c:nonlin} we
present the non-linear simulations and the topography of the turbulent state,
before analyzing in \S\ref{c:clump} the statistics and causes of particle
clumping in more detail. Then \S\ref{c:transport} addresses the transport and
diffusion of particles and angular momentum in the saturated streaming
turbulence. We summarize our results in \S\ref{c:conc}.

\section{Numerical Method} \label{c:numerics}

The dynamical equations and the numerical method are presented in detail in YJ,
but we briefly recapitulate the main points in this section. As a numerical
solver we use the Pencil Code\footnote{See
\url{http://www.nordita.dk/software/pencil-code/}.}. This is a finite
difference code that uses 6th order symmetric spatial derivatives and a 3rd
order Runge-Kutta time integration \citep[see][ for details]{Brandenburg2003}.

\begin{figure}
  \includegraphics[width=8.7cm]{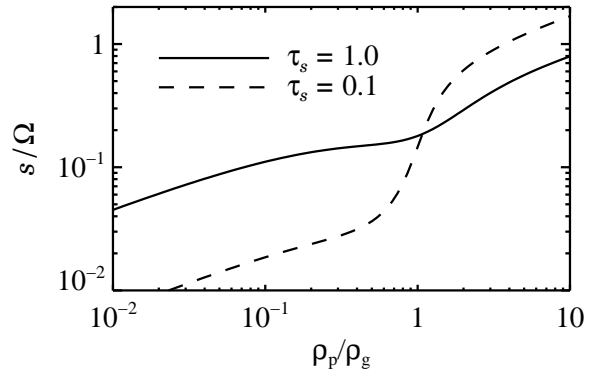}
  \caption{Peak growth rate $s$ of the streaming instability versus the
    solids-to-gas ratio $\epsilon$ for friction times of $\tau_{\rm s}=1.0$
    (solid line) and $\tau_{\rm s}=0.1$ (dashed line). The steep rise in growth
    rate when $\tau_{\rm s}=0.1$ particles cross $\rho_{\rm p}/\rho_{\rm g}=1$
    explains the cavitation in the non-linear run AB (see \Fig{f:epsd_AB}).}
  \label{f:KzInf}
\end{figure}
\begin{deluxetable*}{lllllcr}
  \tablecaption{Run Parameters}
  \tablewidth{0pt}
  \tablehead{
    \colhead{Run} &
    \colhead{$\tau_{\rm s}$} & \colhead{$\epsilon$} &
    \colhead{$L_x \times L_y \times L_z$} &
    \colhead{$N_x \times N_y\times N_z$} &
    \colhead{$N_{\rm p}$} &
    \colhead{$\Delta t$} }
  \startdata
    AA    & $0.1$ & $0.2$ & $\hspace{5pt}4.0 \times \hspace{5pt}4.0 \times\hspace{5pt}
4.0$ &
        $256 \times \,\,\,\,\,\, 1 \times 256$ & --- &
        $2000.0$ \\
    AB    & $0.1$ & $1.0$ & $\hspace{5pt}2.0 \times \hspace{5pt}2.0 \times\hspace{5pt}
2.0$ &
        $256 \times \,\,\,\,\,\, 1 \times 256$ & $1.6\times10^6$ &
        $50.0$ \\
    AC    & $0.1$ & $3.0$ & $\hspace{5pt}2.0 \times \hspace{5pt}2.0 \times\hspace{5pt}
2.0$ &
        $256 \times \,\,\,\,\,\, 1 \times 256$ & $1.6\times10^6$ &
        $50.0$ \\
    BA    & $1.0$ & $0.2$ & $40.0 \times 40.0 \times40.0$ &
        $256 \times \,\,\,\,\,\, 1 \times 256$ & $1.6\times10^6$ &
        $500.0$ \\
    BB    & $1.0$ & $1.0$ & $20.0 \times 20.0 \times20.0$ &
        $256 \times \,\,\,\,\,\, 1 \times 256$ & $1.6\times10^6$ &
        $250.0$ \\
    BC    & $1.0$ & $3.0$ & $20.0 \times 20.0 \times20.0$ &
        $256 \times \,\,\,\,\,\, 1 \times 256$ & $1.6\times10^6$ &
        $250.0$ \\
    AB-3D & $0.1$ & $1.0$ & $\hspace{5pt}2.0 \times \hspace{5pt}2.0 \times\hspace{5pt}
2.0$ &
        $128 \times 128 \times 128$ & $2.0\times10^7$ &
        $35.0$ \\
    BA-3D & $1.0$ & $0.2$ & $40.0 \times 40.0 \times40.0$ &
        $128 \times 128 \times 128$ & $2.0\times10^7$ &
        $300.0$ \\
  \enddata
  \tablecomments{Col. (1): Name of run. Col. (2): Friction time. Col. (3):
    Solids-to-gas ratio. Col. (4): Box size in units of $\eta r$. Col. (5):
    Grid resolution. Col. (6): Number of particles. Col. (7): Total run time in
    units of $\varOmega^{-1}$.}
  \label{t:runs}
\end{deluxetable*}
We model a local patch in a protoplanetary disk with the shearing sheet
approximation \citep[e.g.][]{GoldreichTremaine1978}. A Cartesian coordinate
frame that corotates with the Kepler frequency $\varOmega$ at a distance $r$
from the central star is oriented with $x$, $y$ and $z$ axes pointing radially
outwards, along the orbital direction, and vertically out of the disk (parallel
to the Keplerian angular momentum vector), respectively. We solve the equations
of motion for deviations from Keplerian rotation in an unstratified model,
i.e.\ vertical gravity is ignored. The gas (and not the solids) is subject to
pressure forces, including the global radial pressure gradient, constant in the
local approximation and measured by the dimensionless parameter
\begin{equation}
  \eta \equiv - {\partial P / \partial r \over 2 \rho_{\rm g} \varOmega^2 r} \approx \left({c_{\rm s} \over v_{\rm K}}\right)^2
  \, ,
\end{equation}
where $v_{\rm K} = \varOmega r$ is the Keplerian orbital speed, while $P$,
$\rho_{\rm g}$, and $c_{\rm s}$ are the pressure, density and sound speed of
our isothermal gas. All our simulations use $\eta=0.005$ and $c_{\rm s}/v_{\rm
K}=H/r=0.1$, where $H$ is the gas scale-height. Our results can be applied to
different values of $\eta$ if velocities are scaled by $\eta v_{\rm K}$, the
pressure-supported velocity, and lengths are scales by $\eta r$, the radial
distance between points where Keplerian and pressure-supported velocities are
equal.\footnote{The value of $H/(\eta r)$ ($=20$ in our simulations) changes
with this scaling, but should not affect the results as long as Mach numbers
remain low ($H/r \ll 1$) so that gas compressibility is insignificant.} 

The solids are treated alternatively as a pressureless fluid or as
superparticles that each contain the mass of many actual solid bodies. Solids
and gas mutually interact by frictional drag forces that are linear in the
relative velocity. This models small particles with a friction (or stopping)
time $\tau_{\rm f}$ that is independent of the relative speed between gas and
particles (i.e.\ no turbulent wakes form). The translation from friction time
to the radius of a particle depends only on gas properties and the material
density of the solids. As a rule of thumb the radius of a compact icy particle
in meters is roughly equal to the dimensionless stopping time
\begin{equation} 
 \tau_{\rm s} \equiv \varOmega \tau_{\rm f}
\end{equation}
at Jupiter's location ($r \approx 5\,{\rm AU}$) in standard minimum mass nebula
models \citep{hay81}. 

When solids are treated as numerical particles, we calculate the drag
acceleration by interpolating the gas velocity at the positions of the
particles using a second-order spline fit to the 9 (27) nearest grid points
that surround a given particle for 2-D (3-D) grids. To conserve momentum we
assign the drag force on each single particle back to the gas using a
Triangular Shaped Cloud (TSC) scheme \citep{HockneyEastwood1981}. This
smoothing of the particles' influence helps overcome shot noise, and should not
be seen as an SPH (smoothed particle hydrodynamics) approach since our
particles carry no hydrodynamic properties. We showed in YJ that 1 particle per
grid point is enough to resolve the linear growth of the streaming instability,
but to better handle Poisson fluctuations for a wider range of densities, we
generally use 25 particles per grid point in the non-linear simulations.

An equilibrium solution to the coupled equations of motion of the gas and the
solids was found by \citet[][ hereafter referred to as NSH]{Nakagawa+etal1986}
where drag balances the radial pressure gradient and Coriolis forces.  YG found
that this equilibrium triggered a linear drag instability. The peak growth rate
of this streaming instability is shown in \Fig{f:KzInf} as a function of the
solids-to-gas ratio $\epsilon$ for $\tau_{\rm s}$ of 0.1 and 1.0. See Figs.\ 1
and 2 of YJ (and the accompanying text) for the dependence of growth rates on
wavenumber.

\section{Non-Linear Evolution to Turbulence}\label{c:nonlin}

\begin{figure*}
  \includegraphics{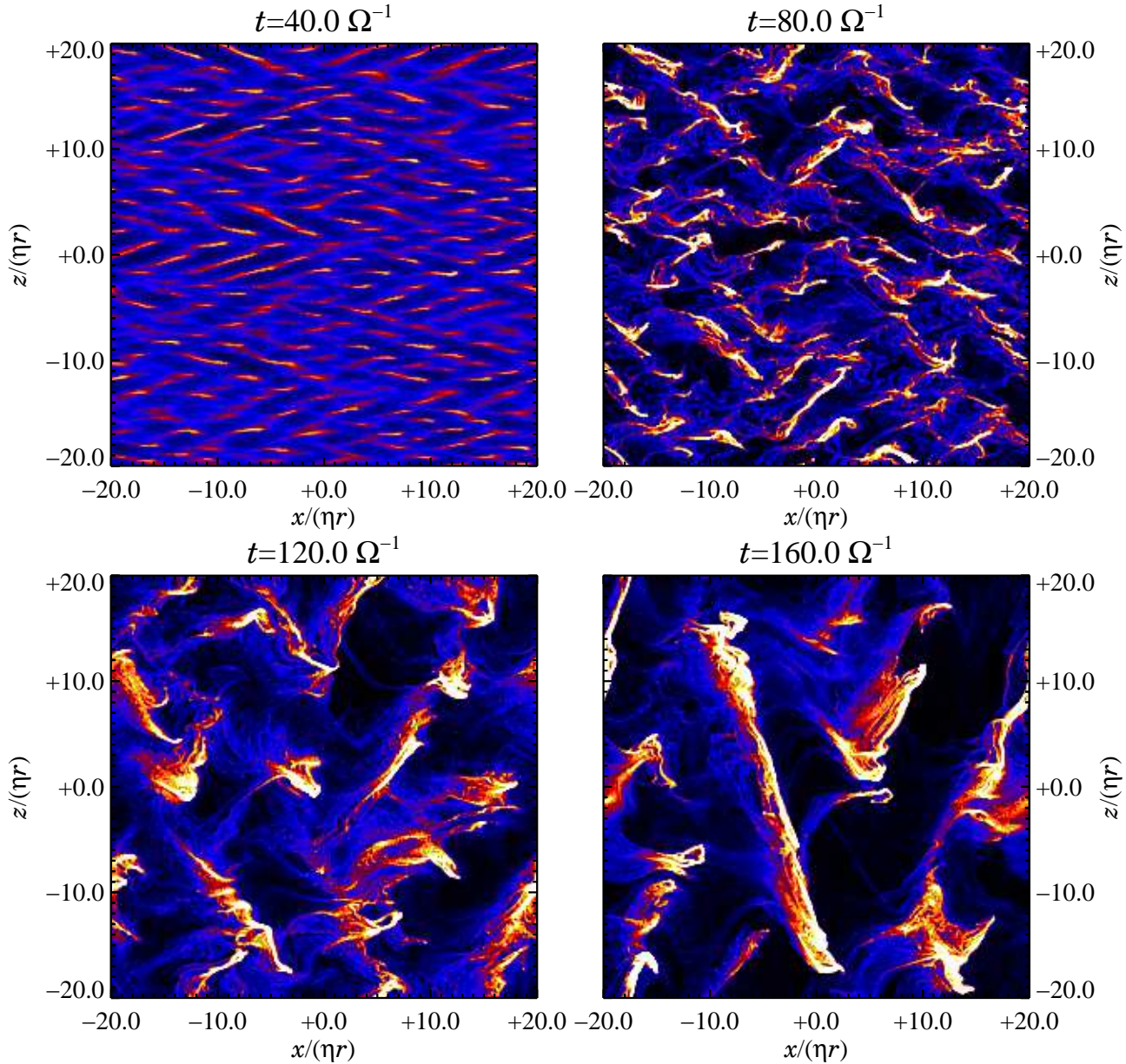}
  \caption{Particle density snapshots for run BA with friction time
    $\tau_{\rm s}=1.0$ and a solids-to-gas ratio of $\epsilon=0.2$. Particle
    densities increase from black (zero density) to bright yellow/white
    (solids-to-gas of unity or higher). The evident linear wavelength in the
    first frame results from the streaming instability feeding off the drift of
    the particles through the gas. Subsequent frames document a surprising
    consequence of the self-consistently generated turbulence: the non-linear
    cascade of dense particle clumps into larger filaments.}
  \label{f:epsd_BA}
\end{figure*}
With confidence from YJ that the code solves correctly for the linear growth of
the streaming instability, we turn our focus to the non-linear evolution into
turbulence. We generically refer to the non-linear states of our runs as
``turbulent," because they contain stochastic fluctuations that diffuse
material and momentum. Some cases (the gas-dominated AA and BA runs, see below)
appear more wave-like with peaks in the spatial and temporal Fourier spectra
(as we will see in \Fig{f:power_rhop}). However, even these runs exhibit
diffusion and stochastic fluctuations on a range of scales, so we also label
them as turbulent. This section describes the simulation parameters and main
results for marginal vs.\ tighter coupling. More detailed analyses of the
turbulent state follow in later sections.

\subsection{Run Parameters and Initialization}

The parameters of the different non-linear simulations are listed in
\Tab{t:runs}. We consider two particle sizes, represented as friction times:
tightly coupled solids with $\tau_{\rm s}=0.1$ (those runs are labeled A*,
where * represents a solids-to-gas ratio label) and larger, more loosely
coupled particles with $\tau_{\rm s}=1.0$ (labeled B*). Three values of the
solids-to-gas mass ratio, $\epsilon=0.2,1.0,3.0$ (labeled *A, *B, *C,
respectively) are considered for each friction time. For instance model AB uses
$\tau_{\rm s} = 0.1$ and $\epsilon = 1.0$. The chosen particle abundances are
well above the Solar composition of $\epsilon \sim 0.01$, but can very well be
achieved in a sedimented mid-plane layer of solids, depending on turbulent
diffusion and various particle enrichment or gas depletion mechanisms.

\begin{figure*}
  \includegraphics{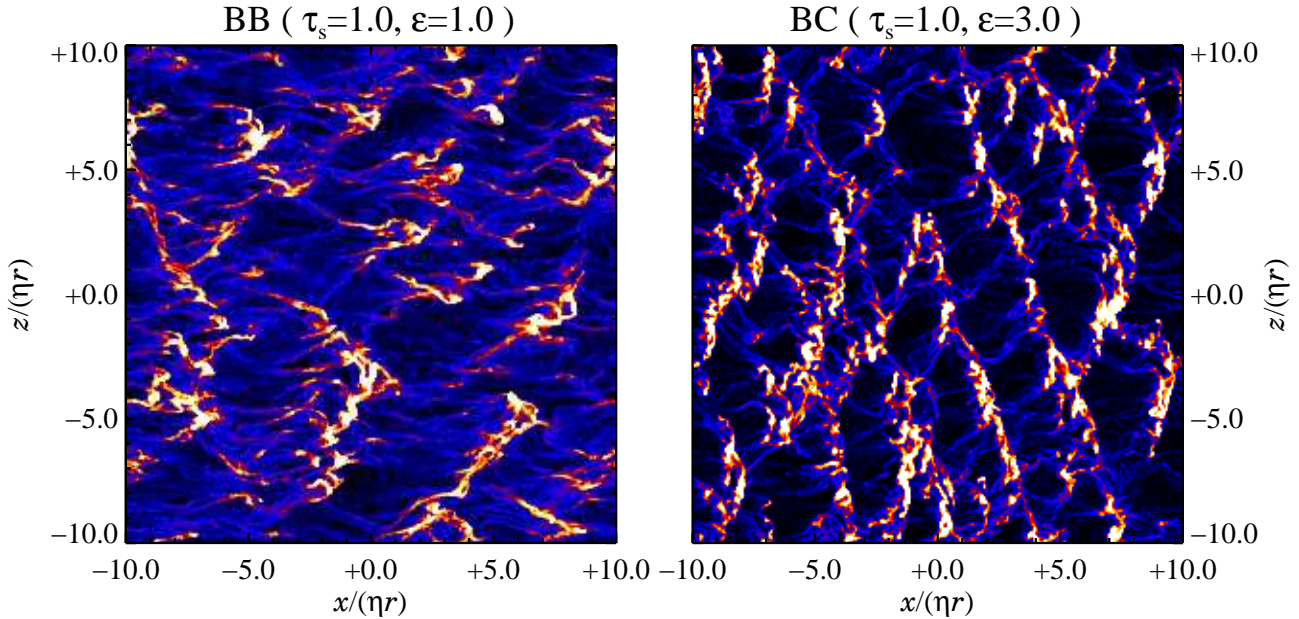}
  \caption{The saturated state of runs BB and BC (both at a time of
    $t=100\varOmega^{-1}$). The range of the solids-to-gas ratio is from 0-5 in
    the left plot and from 0-15 in the right plot, giving both the same
    relative scale for particle overdensities as \Fig{f:epsd_BA}. The tendency
    for dense clumps to lean against the radial drift flow is evident.}
  \label{f:epsd_BB+BC}
\end{figure*}
The size of the simulation box was chosen in all cases such that the most
unstable radial wavelengths are resolved with at least 8 grid points. Two of
the runs (labeled AB-3D, BA-3D) were fully 3-D, all others were 2.5-D
simulations of the radial-vertical plane with all three velocity components,
consistent with the linear analysis of YG. Fully periodic boundary conditions
were used for the 2.5-D runs, while the 3-D simulations impose a shear-periodic
boundary condition in the radial direction (see \S3.2.1 of YJ). Particles are
initially placed randomly throughout the simulation box. This ``warm start''
gives a white noise power spectrum with scale-independent Fourier amplitudes of
$\tilde\rho_{{\rm p}}(\vc{k})/\langle\rho_{{\rm p}}\rangle \sim1/\sqrt{N_{\rm
p}}$ in the particle density. The noise serves as a seed for streaming
instabilities. The velocities of gas and solids are initially set to the
equilibrium values of NSH.

\subsection{Marginally Coupled Boulders}\label{s:boulders}

Many drag force phenomena are most prominent for marginally coupled, $\tau_{\rm
s}=1$, particles, corresponding to approximately meter-sized boulders at $r
\approx 5$ AU in the solar nebula. Streaming instabilities are no exception,
with fast linear growth\footnote{Somewhat paradoxically, tight coupling gives
faster growth in the particle-dominated regime, but on smaller scales.} and
significant particle clumping in this regime. \Fig{f:epsd_BA} shows four
snapshots of the evolution of the streaming instability into turbulence for run
BA ($\tau_{\rm s}=1.0$ $\epsilon=0.2$). The initial growth is dominated by the
fastest linear modes (first frame of \Fig{f:epsd_BA}), consistent with the
maximum analytic growth rate, $s \approx 0.1 \varOmega$ for $k_x \eta r \approx
1$ (see \Fig{f:KzInf} and also Figs. 1 \& 2 of YJ).

A non-linearly fluctuating, i.e.\ turbulent, state is reached after some 80
local shear times (second frame of \Fig{f:epsd_BA}). Solids become concentrated
in a few massive clumps surrounded by an ocean of lower density material.
Radial drift speeds are lower in such dense regions (we discuss the reduced
radial drift further in \S\ref{s:radialdrift}). Solid particles are eventually
lost downstream from the clumps into the voids, where the radial drift is
faster, until they fall into another dense particle clump. Over a time-scale of
more than 100 shear times (third and fourth frame of \Fig{f:epsd_BA}) this
leads to an upward cascade of the density structure into extended filaments
(actually rolls and sheets if we extend into the symmetric azimuthal
dimension). The filaments are predominantly aligned in the vertical direction,
which maximizes their ability to intercept particles, but are slightly tilted
radially in alternating directions. Strong bulk motions are exhibited by the
filaments along their long axis. This helps them stay upwind (motions are in
the $+z,+x$ or $-z,+x$ directions), and leads to their disruption in several
orbital times when alternately aligned filaments collide.\footnote{The behavior
described is best seen in a movie of run AB which can be downloaded from\\
\url{http://www.mpia.de/homes/johansen/research\_en.php}.} The bulk motion also
leads to efficient mixing of particles, especially in the vertical direction
(see \S\ref{s:turbdiff}). The extended filaments are closely related to the
long-lived vertically-oscillating clumps seen in 2-D simulations of the
Kelvin-Helmholtz instability with $\tau_{\rm s} = 1.0$ particles \citep[see
Fig.\ 8 of][]{jhk06}.

\begin{figure}
  \includegraphics{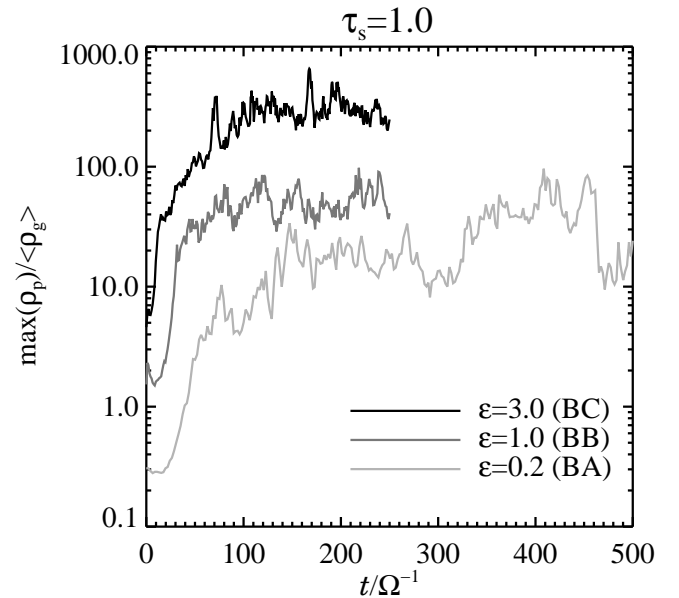}
  \caption{Maximum bulk density of solids, in units of the average gas density
    in the box, as a function of time for the three marginally coupled runs.
    The maximum density is generally around two orders of magnitude higher than
    the mean bulk density of the solids. The particle density has been assigned
    to the mesh using the TSC scheme.}
  \label{f:rhopmax_t_t1.0}
\end{figure}

The $\tau_{\rm s} = 1.0$ runs with larger $\epsilon$ values (BB and BC) evolve
similarly to run BA, but with a less pronounced cascade to larger scales (the
saturated states of those two runs are shown in \Fig{f:epsd_BB+BC}).
\Fig{f:rhopmax_t_t1.0} shows the evolution of maximum particle density
(assigned with the TSC scheme to the grid) versus time for all three runs. The
non-linear state is characterized by density peaks $100$ times (or more) above
the average particle density. Run BA has a longer run time, not only because
the gas-dominated case is more astrophysically interesting, but also because it
took longer to reach a saturated state. \Fig{f:rhopmax_t_t1.0} shows signs of
secular growth in densities over the entire $\Delta t=500\varOmega^{-1}$.  Even
longer runs would better characterize long term fluctuations in the saturated
state, but such computational resources would probably be better spent on a
more realistic model with vertical gravity.

\begin{deluxetable*}{lllllllllll}
  \tablecaption{Flow Properties}
  \tablewidth{0pt}
  \tablehead{
    \colhead{Run} &
    \colhead{$\tau_{\rm s}$} & \colhead{$\epsilon$} &
    \colhead{${\rm Ma}_x$} &
    \colhead{${\rm Ma}_y$} &
    \colhead{${\rm Ma}_z$} &
    \colhead{$\overline{v_x}$} &
    \colhead{$\overline{v_x}^{\rm (NSH)}$}
    }
  \startdata
    AA    & $0.1$ & $0.2$ & $5.7\times10^{-4}$ & $1.1\times10^{-3}$ &
            $3.4\times10^{-3}$ & $-0.138$ & $-0.138$ \\
    AB    & $0.1$ & $1.0$ & $1.2\times10^{-2}$ & $6.1\times10^{-3}$ &
            $8.5\times10^{-3}$ & $-0.108$ & $-0.050$ \\
    AC    & $0.1$ & $3.0$ & $8.7\times10^{-3}$ & $4.5\times10^{-3}$ &
            $6.4\times10^{-3}$ & $-0.035$ & $-0.012$ \\
    BA    & $1.0$ & $0.2$ & $1.2\times10^{-2}$ & $1.8\times10^{-2}$ &
            $4.0\times10^{-2}$ & $-0.520$ & $-0.820$ \\
    BB    & $1.0$ & $1.0$ & $9.3\times10^{-3}$ & $1.1\times10^{-2}$ &
            $9.2\times10^{-3}$ & $-0.341$ & $-0.400$ \\
    BC    & $1.0$ & $3.0$ & $8.9\times10^{-3}$ & $1.3\times10^{-2}$ &
            $1.1\times10^{-2}$ & $-0.118$ & $-0.118$ \\
    AB-3D & $0.1$ & $1.0$ & $5.3\times10^{-3}$ & $3.4\times10^{-3}$ &
            $2.7\times10^{-3}$ & $-0.064$ & $-0.050$ \\
    BA-3D & $1.0$ & $0.2$ & $1.2\times10^{-2}$ & $1.7\times10^{-2}$ &
            $3.3\times10^{-2}$ & $-0.545$ & $-0.820$
  \enddata
  \tablecomments{Col. (1): Name of run. Col. (2): Friction time. Col. (3):
    Solids-to-gas ratio. Col. (4)-(6): Turbulent Mach number of the gas. Col.
    (7): Mean radial particle velocity in units of $\eta v_{\rm K}$. Col. (8):
    Mean radial particle velocity in NSH state.}
  \label{t:flow}
\end{deluxetable*}
\begin{figure*}
  \includegraphics{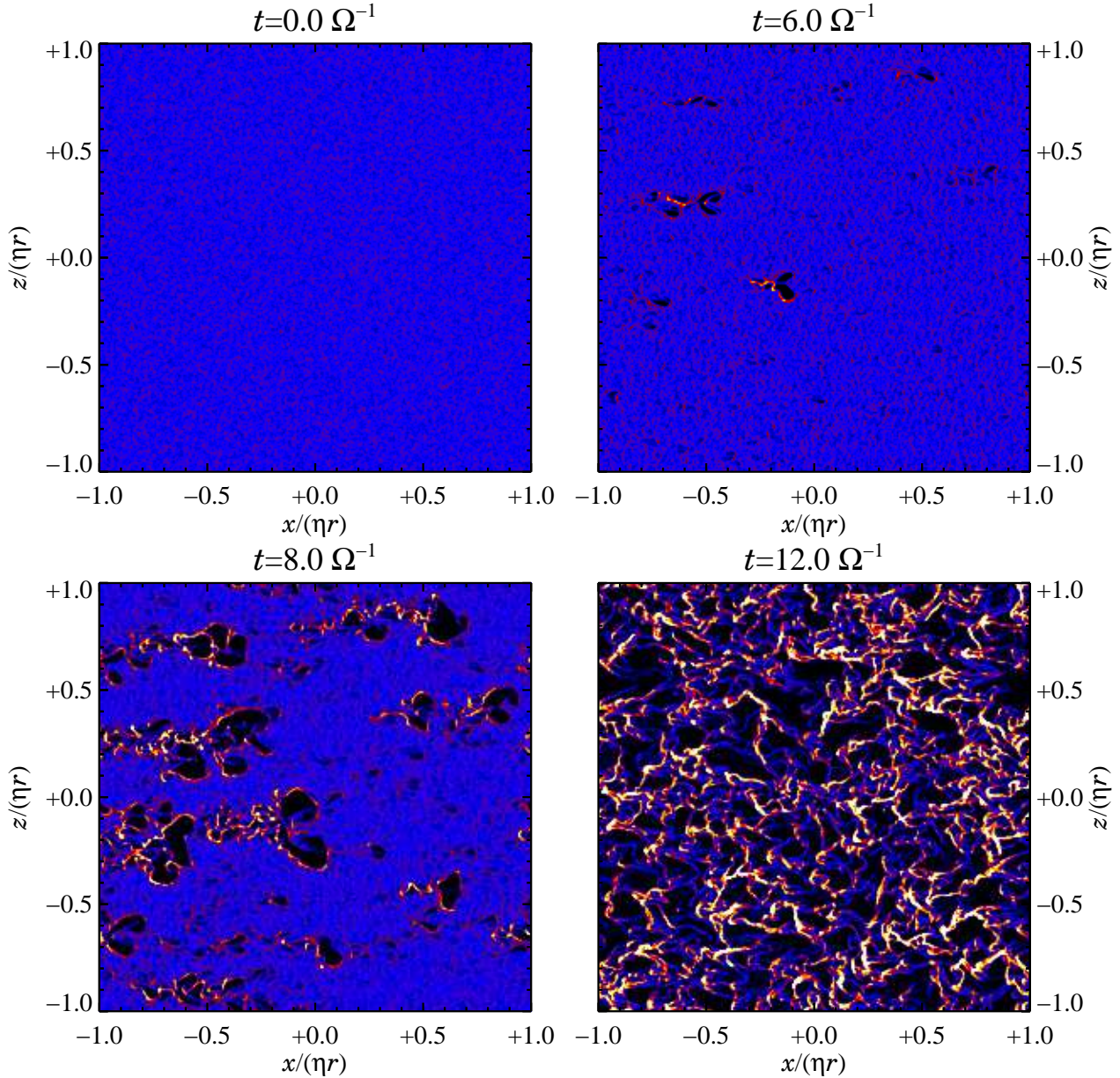}
  \caption{The onset of streaming turbulence for run AB ($\tau_{\rm s}=0.1$,
    $\epsilon=1.0$). The plots show color coded particle density (black is zero
    particle density, bright is a solids-to-gas ratio of 5 or higher). The
    first frame shows only the initial Poisson noise. After around one orbital
    period small voids form. The inner edges of the cavities are loaded with
    particles that can fall rapidly through the voids in the absence of any
    collective drag force effects there. The voids rapidly expand, and a
    self-sustained turbulent state sets in after around 2 orbits. This atypical
    onset of turbulence is caused by an increased growth rate of the streaming
    instability in slightly overdense grid cells (see \Fig{f:KzInf}).}
  \label{f:epsd_AB}
\end{figure*}
\Tab{t:flow} lists turbulent Mach numbers of the gas flow (after subtracting
the mean flow, see \S\ref{s:GasNoClump}). The anisotropic turbulence of run BA,
with stronger fluctuations in the vertical direction, is clear. Turbulence is
more isotropic in the other marginally coupled runs.

\subsection{Tightly Coupled Rocks}\label{s:tight}

Simulations with shorter friction times are more costly because the shorter
unstable length-scales (Figs.\ 1 \& 2 of YJ) impose more stringent Courant
criteria. The effort was nevertheless rewarded with a qualitatively very
different behavior for the $\tau_s = 0.1$ runs. These particles correspond to
solid rocks of approximately 10 cm size at $r=5$ AU in the solar nebula.

\subsubsection{$\tau_s = 0.1$, $\epsilon = 1.0$: Cavitation}

\Fig{f:epsd_AB} shows four snapshots of the particle density for run AB
(friction time $\tau_{\rm s}=0.1$, solids-to-gas ratio $\epsilon=1.0$). The
first frame displays the initial Poisson noise. In contrast to run BA (see the
first frame of \Fig{f:epsd_BA}), we do not see the smooth growth of linear
waves over ten or more of orbital times (an expectation which follows from the
peak growth rate $s \approx 0.15 \varOmega$). Instead a few voids with dense
inner rims appear by $t = 6 \varOmega^{-1}$ (second frame). The cavities expand
rapidly (third frame), leading to a fully turbulent state after only two
orbits, i.e.\ $t \approx 12 \varOmega^{-1}$ (fourth frame).

The effect of Poisson fluctuations on the linear growth properties in
\Fig{f:KzInf} largely explains the surprisingly rapid and non-uniform growth.
The $\epsilon = 1.0$ state lies amid a steep rise in growth rates from the
gas-dominated to particle-dominated regimes. Specifically the growth time for
$\epsilon = 1.0$, $t_{\rm grow} \equiv 1/s = 6.8 \varOmega^{-1}$, is halved for
a modest increase in the solids-to-gas ratio to $\epsilon = 1.25$. This
enhanced growth applies in locally overdense regions. Poisson fluctuations from
assigning $N_{\rm p} = 1.6 \times 10^{6}$ particles to $N_{\rm b} = 256^2$ bins
generate density fluctuations with a standard deviation of $\delta_{\rm p}
\simeq \sqrt{N_{\rm b}/N_{\rm p}} \approx 0.2$. The TSC assignment smooths
these fluctuations somewhat, but random overdensities $25 \%$ or greater still
exist in over 1000 cells ($1.6 \%$ of the total). Since a region with $\epsilon
\approx 1.25$ is already non-linear after two e-foldings (consistent with the
observed $t \approx 6 \varOmega^{-1}$), enhanced growth in overdense regions
plausibly explains the growth of cavities. 

\begin{figure}
  \includegraphics{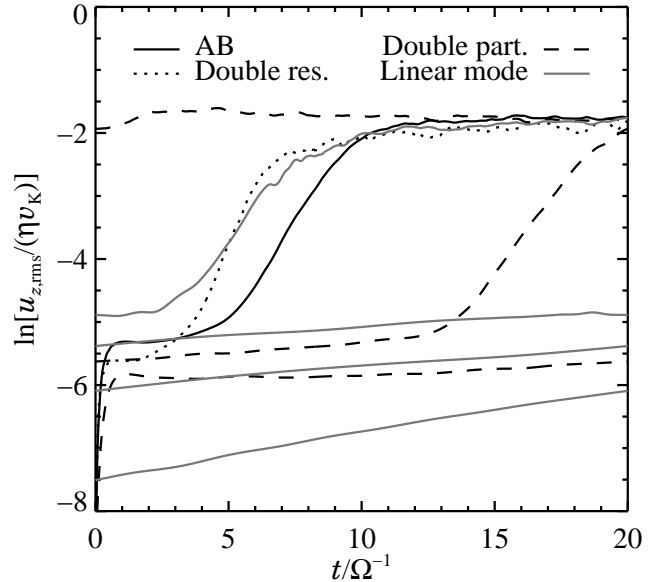}
  \caption{Growth of the streaming instability, as measured by the amplitude of
    vertical velocity fluctuations, for different numerical approaches to run
    AB. Only a seeded linear mode (gray line) grows with the expected growth
    rate of $s \approx 0.1\varOmega$ (some curves have been wrapped around
    several times to allow simulations to run further). Fast growing cavities
    occur both at double resolution (dotted line) and with twice as many
    particles per grid cell (dashed line). Since cavities are triggered by
    Poisson density fluctuations, explosive growth is delayed with twice as
    many particles. The saturated state is the same for all cases.}
  \label{f:cavity_growth}
\end{figure}
We confirm this physical explanation for the cavities by running five
variations to AB: (1) doubling the spatial resolution, (2) doubling the number
of particles per grid cell, (3) seeding a linear mode (an eigenvector) with the
``cold start" algorithm used for the linear tests in YJ, (4) the same linear
mode, but the particle density distribution is seeded randomly (and thus
dominated by Poisson fluctuations), and (5) quadratic polynomial instead of
spline interpolation (see Appendix A of YJ). The growth of the root-mean-square
of the vertical gas velocity for the first three variations is shown in
\Fig{f:cavity_growth} along with the original run BA. Variation (1) [and also
(4) and (5), not shown] give essentially the same behavior run BA, which
eliminates obvious numerical effects (grid resolution and interpolation scheme)
as the source of cavities.\footnote{Variation (5) is interesting because the
Poisson density fluctuations dominate the carefully seeded velocities.}
Doubling the particle number, variation (2), delays the onset of cavitation, as
expected with lower amplitude Poisson fluctuations. Variation (3) suppresses
all Poission noise, and the ``cold" linear mode grows at the analytic rate,
$s\approx0.1\varOmega$, until non-linear effects finally dominate after $t = 60
\varOmega^{-1}$. Perhaps most importantly, all approaches lead to the same
saturated state, despite markedly different routes to turbulence. This speaks
to the robustness, not just of transient cavitation, but of all the non-linear
results.
 
We also investigated the velocity structure at the onset of cavitation.
Quadrupolar structures (most prominent in the vertical velocity) appear as
isolated modes of the streaming instability. The length scale of the
quadrupolar distortions did not vary upon doubling the grid resolution (with a
fixed number of particles per grid cell).

\begin{figure*}
  \includegraphics{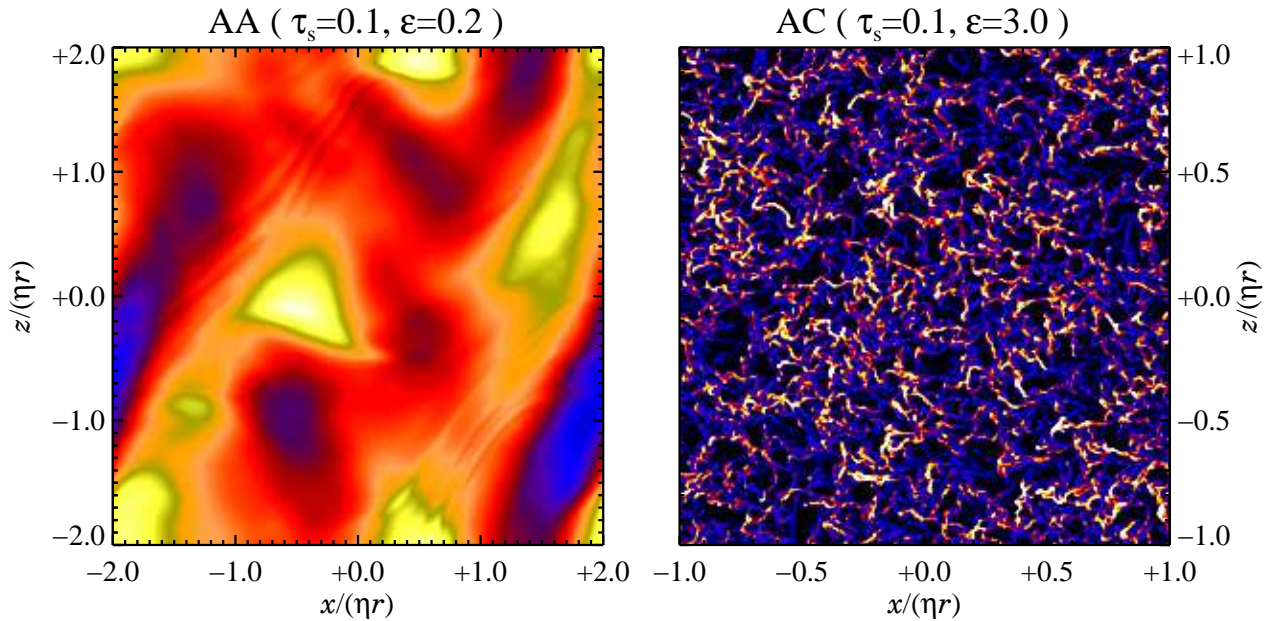}
  \caption{The saturated states of run AA (at $t=1000\varOmega^{-1}$) and run AC
    (at $t=50\varOmega^{-1}$). The range in solids-to-gas ratio is 0.15-0.25 in
    the left plot and 0-15 in the right plot. Run AA (calculated with the two
    fluid code, see \S\ref{s:AA} for explanation) is dominated by oscillatory
    motion of slightly overdense clumps. The turbulent state of run AC is very
    much like run AB, but at smaller scales. Also the non-linear state of run
    AC develops simultaneously throughout the grid, unlike the cavitation of
    run AB.}
  \label{f:epsd_AA+AC}
\end{figure*}

Fortunately, our Poisson noise hypothesis does not predict cavitation where it
should not occur. Run AC ($\tau_s = 0.1$, $\ep = 3.0$) has a fast linear growth
rate with a relatively weak dependence on the local value of $\ep$ (see
\Fig{f:KzInf}). Accordingly non-linear fluctuations appear uniformly throughout
the grid in run AC, instead of cavitating first in a few spots. The saturated
state of run AC (shown in the right panel of \Fig{f:epsd_AA+AC}) is similar to
run AB, but with smaller scale fluctuations. Like run AB, the marginally
coupled run BB has equal densities of particles and gas, but with $\tau_{\rm s}
= 1.0$ the rise in growth rates across $\epsilon = 1$ was much smoother (see
\Fig{f:KzInf}).\footnote{Fig. 3 of YG confirms this trend, showing that the
transition is yet sharper for $\tau_s =0.01$ ``pebbles."} Since the effect of
Poisson fluctuations is weak (an overdensity of $25 \%$ only cuts the growth
time by $12\%$ for $\tau_s = 1.0$ instead of halving it for $\tau_s = 0.1$) run
BB displays orderly growth of the dominant linear modes. 

\subsubsection{$\tau_s = 0.1$, $\epsilon = 0.2$: Weak Overdensities}\label{s:AA}
 
\begin{figure}
  \includegraphics{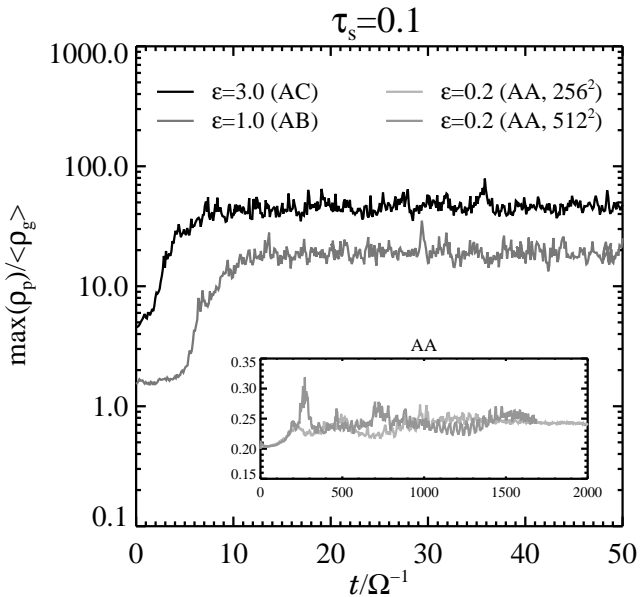}
  \caption{Maximum bulk density of solids for the three tightly coupled runs
    (AB, AC and in the insert AA at two different resolutions). The maximum
    density is around an order of magnitude higher than the average for the
    $\epsilon=1.0,3.0$ runs, whereas the turbulent state of the $\epsilon=0.2$
    run only experiences very mild relative overdensities of around 20\% at
    both $256^2$ and $512^2$ mesh resolution.}
  \label{f:rhopmax_t_t0.1}
\end{figure}
\Fig{f:rhopmax_t_t0.1} plots the maximum particle density versus time for the
three tightly coupled simulations. The streaming instability produces particle
overdensities of 20 or more in runs AB and AC. However the gas-dominated case
AA has a qualitatively different behavior. Growth saturates (see left panel of
\Fig{f:epsd_AA+AC}) in a few growth times, $t_{\rm grow} = 1/s \approx 42
\varOmega^{-1}$, as expected. However the particle overdensities are very mild,
only 20\% on average (see inset of \Fig{f:rhopmax_t_t0.1}). To test for
convergence we ran the simulation at both $256^2$ and $512^2$ grid points, but
the qualitative evolution of maximum bulk density of solids is unchanged (after
a small initial peak in the $512^2$ run).

We emphasize that run AA was performed with the two-fluid code, not the
particle-fluid approach used in the other simulations. This choice was
necessitated by computational cost of long growth times with short wavelengths
that restrict the code to small time steps. It is tempting to suspect that the
weak overdensities in AA are a consequence of the two-fluid approach. However
the limitation of the pressureless fluid model of solids is that density
gradients steepen and shock, causing numerical instabilities, \emph{not} that
they are stably smoothed. To confirm this we ran two-fluid simulations of case
AB and obtained the expected (from the particle-fluid run) growth of particle
density until the code crashed after the growth of non-linear overdensities. By
contrast AA simply never generates large density fluctuations, apparently since
drag feedback on the gas is too weak in the small particle, gas-dominated
regime. 

\subsection{3-D Simulations}\label{c:3d}

\begin{figure*}
  \includegraphics[width=8.0cm]{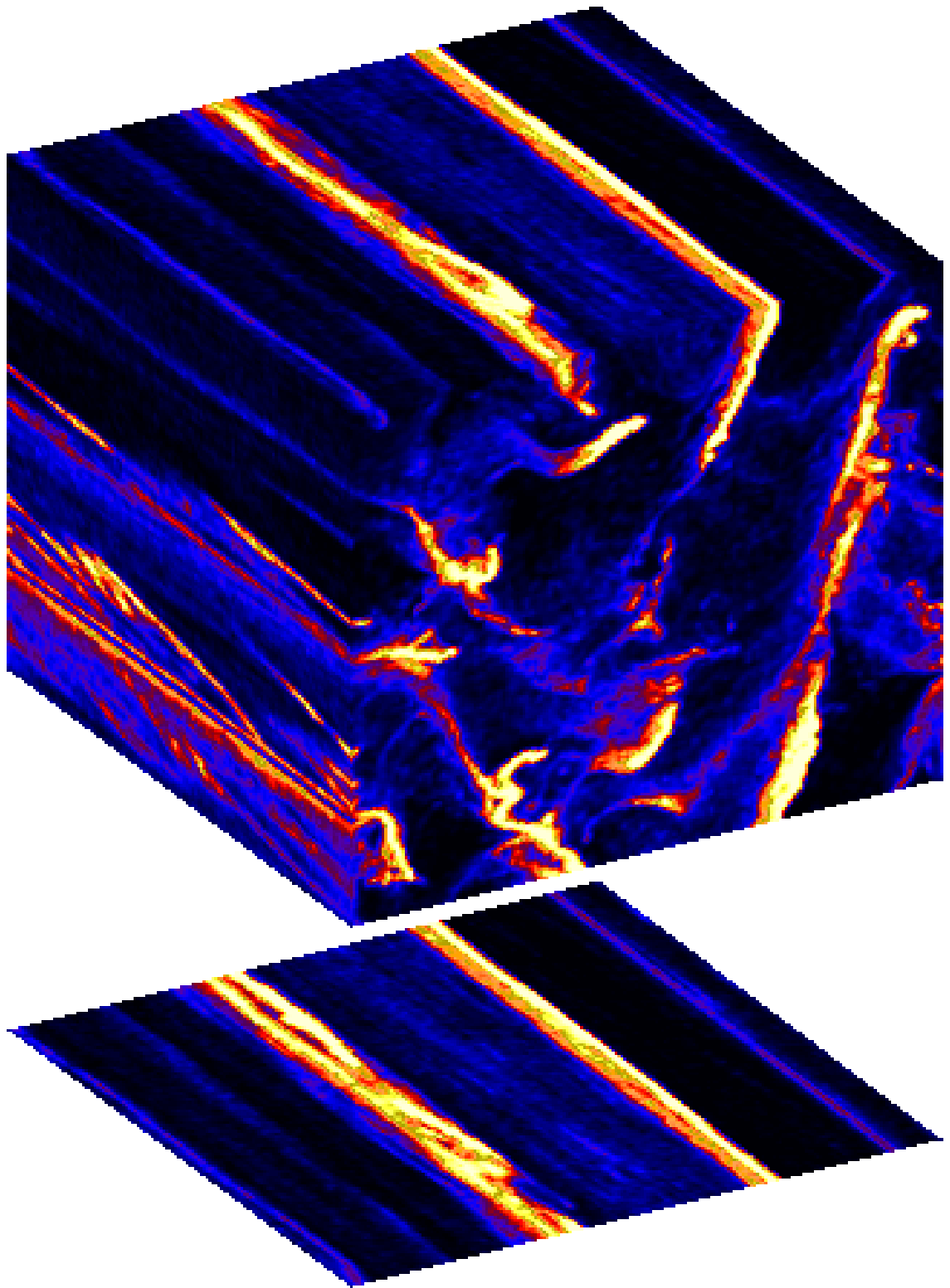}
  \includegraphics[width=8.0cm]{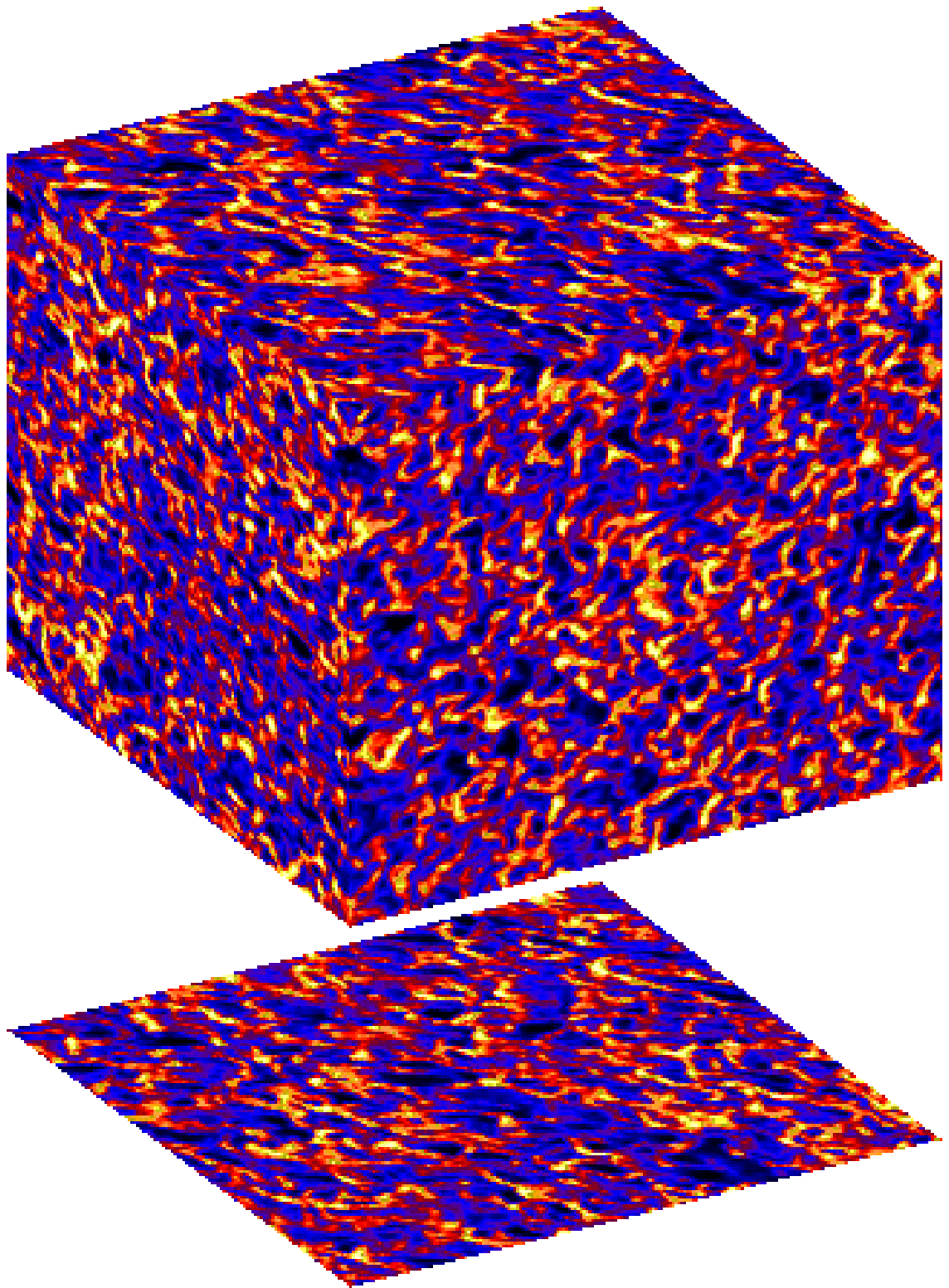}
  \caption{Saturated streaming turbulence for run BA-3D ($\epsilon=0.2$,
    $\tau_{\rm s}=1.0$, left) and run AB-3D ($\epsilon=1.0$, $\tau_{\rm
    s}=0.1$, right). The boxes are oriented with the radial $x$-axis to the
    right and slightly up, the azimuthal $y$-axis to the left and up, and the
    vertical $z$-axis directly up. The contours show the particle density at
    the sides of the simulation box after the streaming turbulence has
    saturated. The axisymmetry of the marginally coupled particles witnesses the
    smearing effect of Keplerian shear on the relatively long-lived clumps. The
    tightly coupled particles drive rapid fluctuations that develop fully
    non-axisymmetric density patterns.}
  \label{f:streaming3D}
\end{figure*}
We have also performed full 3-D simulations of the streaming instability with
$128^3$ grid points and $N_{\rm p}=2 \times 10^7$ particles. The linear
analysis of YG assumed axisymmetry, partly for simplicity, but also because
modes that grow slowly (with $s < \varOmega$) will be sheared into rings in any
event. Computer simulations are needed to determine whether the saturated state
remains azimuthally symmetric in 3-D and how the presence of the azimuthal
direction affects turbulent properties. Note that even axisymmetric linear
instabilities can give rise to non-axisymmetric parasitic instabilities
\citep[e.g.\ Kelvin-Helmholtz instabilities that feed off the channel flow of
the magnetorotational instability, see][]{GoodmanXu1994}.

\Fig{f:streaming3D} shows the particle density for runs BA-3D and AB-3D in a
saturated state. The marginally coupled case (BA-3D) maintains a high degree of
axisymmetry. The radial-vertical plane shows the cascade into sheets similar to
the 2.5-D case (as seen in \Fig{f:epsd_BA}). The quantitative analysis of
turbulent properties (see Tables \ref{t:flow} and \ref{t:turbulence}) confirms
that BA-3D is very similar to the 2.5-D case. The ability to maintain azimuthal
symmetry suggests (as we will confirm in \S\ref{c:tcorr}) that particles reside
in clumps for longer than an orbital time, so that clumps become azimuthally
elongated by radial shear. Notice that the clump lifetime is not so long that
structures appear perfectly axisymmetric. 

\begin{deluxetable*}{lrrrrrrr}
  \tablecaption{Turbulent Transport}
  \tablewidth{0pt}
  \tablehead{
    \colhead{Run} &
    \colhead{$\tau_{\rm s}$} & \colhead{$\epsilon$} &
    \colhead{$D_x$} &
    \colhead{$D_z$} &
    \colhead{$\mathcal{F}_{\mathcal{L},x}^{\rm (turb)}$} &
    \colhead{$\mathcal{F}_{\mathcal{L},x}^{\rm (NSH)}$}
    }
  \startdata
    AA    & $0.1$ & $0.2$ &
            $(1.4\pm6.2)\times10^{-7}$ & $(6.0\pm262)\times10^{-7}$ &
            $-2.2\times10^{-8}$ & $-4.8\times10^{-7}$ \\
    AB    & $0.1$ & $1.0$ &
            $(4.4\pm0.4)\times10^{-5}$ & $(2.9\pm0.5)\times10^{-5}$ &
            $-6.1\times10^{-5}$ & $-3.1\times10^{-7}$ \\
    AC    & $0.1$ & $3.0$ &
            $(2.0\pm0.2)\times10^{-5}$ & $(1.8\pm0.2)\times10^{-5}$ &
            $-6.0\times10^{-5}$ & $-5.8\times10^{-8}$ \\
    BA    & $1.0$ & $0.2$ &
            $(2.2\pm0.6)\times10^{-3}$ & $(1.5\pm0.8)\times10^{-2}$ &
            $6.7\times10^{-5}$ & $-1.7\times10^{-4}$ \\
    BB    & $1.0$ & $1.0$ &
            $(7.6\pm0.7)\times10^{-4}$ & $(1.7\pm0.4)\times10^{-4}$ &
            $-4.0\times10^{-5}$ & $-2.0\times10^{-4}$ \\
    BC    & $1.0$ & $3.0$ &
            $(2.8\pm0.2)\times10^{-4}$ & $(6.2\pm0.9)\times10^{-4}$ &
            $-1.5\times10^{-4}$ & $-5.2\times10^{-5}$ \\
    AB-3D & $0.1$ & $1.0$ &
            $(1.6\pm0.2)\times10^{-5}$ & $(2.7\pm0.1)\times10^{-6}$ &
            $-1.5\times10^{-5}$ & $-3.1\times10^{-7}$ \\
    BA-3D & $1.0$ & $0.2$ &
            $(2.0\pm0.3)\times10^{-3}$ & $(8.2\pm2.5)\times10^{-3}$ &
            $6.0\times10^{-5}$ & $-1.7\times10^{-4}$ \\
  \enddata
  \tablecomments{Col. (1): Name of run. Col. (2): Friction time. Col. (3):
    Solids-to-gas ratio. Col. (4)-(5): Turbulent diffusion coefficient in units
    of $c_{\rm s}^2 \varOmega^{-1}$ (interval indicates one standard deviation
    in each direction). Col. (6): Radial flux of azimuthal momentum relative
    to NSH state. Col. (7): Radial flux of azimuthal momentum in NSH state.
    All quantities are normalized with standard combinations of $\varOmega$,
    $c_{\rm s}$ and $\rho_{\rm g}$.}
  \label{t:turbulence}
\end{deluxetable*}
The tightly coupled case (AB-3D) on the other hand evolves completely
non-axisymmetrically. Indeed the correlation time of the clumps is short enough
that they are not significantly elongated by Keplerian shear. Similar to the
2.5-D case, cavities (now fully 3-D and non-axisymmetric) developed out of the
initial Poisson noise in run AB-3D. The saturated state appears to have less
pronounced clumps than run AB (the fourth panel of \Fig{f:epsd_AB}).  Tables
\ref{t:flow} and \ref{t:turbulence} show that the 3-D turbulence indeed has
lower velocities (Mach numbers) and weaker diffusion, particularly in the
vertical direction. It is to be expected that turbulent properties in the 3-D
runs change more for the case that is non-axisymmetric (AB-3D) than the case
that remains axisymmetric (BA-3D) and was already capturing the relevant
physics.

\begin{figure}
  \includegraphics{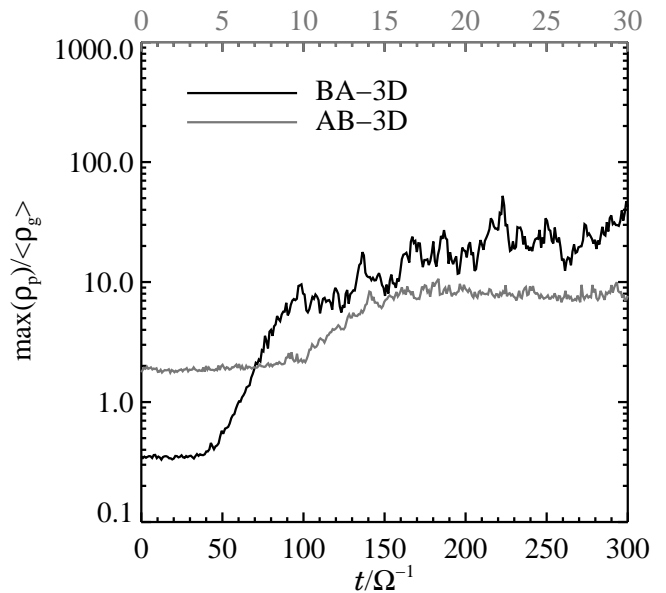}
  \caption{Evolution of the maximum bulk density of solids in the two 3-D
    simulations (notice the different time axes). The density of AB-3D is
    somewhat lower than in the 2.5-D case (\Fig{f:rhopmax_t_t0.1}), whereas the
    marginally coupled BA-3D shows good agreement with run BA in
    \Fig{f:rhopmax_t_t1.0}.}
  \label{f:rhopmax_t_3D}
\end{figure}
The peak particle densities for the two 3-D runs are shown in
\Fig{f:rhopmax_t_3D}. Compared to the density evolution of the 2.5-D runs
(\Figs{f:rhopmax_t_t1.0}{f:rhopmax_t_t0.1}) it is evident that BA-3D agrees
well with BA, whereas AB-3D achieves a somewhat lower maximum density than AB
does.  We will focus most of our analysis on the 2.5-D runs because we could
conduct a more systematic study of parameter space at higher spatial
resolution. The 3-D runs presented here support this choice by giving
qualitatively (and for BA, fairly quantitatively) similar results to the 2.5-D
runs.

\section{Particle Concentration}\label{c:clump}

The ability of drag forces to concentrate particles via the non-linear
evolution of the streaming instability is now analyzed in detail. This
fundamentally important process could alter the collisional evolution of the
size spectrum of particles, leading to an enhanced growth of the average
particle radius, or even trigger gravitational instabilities in the solid
component of protoplanetary disks.

\subsection{Gas Does Not Clump}\label{s:GasNoClump}

We emphasize that gas densities remain nearly constant, despite non-linear
particle overdensities in streaming turbulence.  Gas overdensities are
$\lesssim 1\%$ in all runs. This validates our use of a constant stopping time,
$\tau_{\rm f}$ (which would otherwise vary with gas density in the Epstein
regime). We note that the linear analysis of YG assumed a perfectly
incompressible gas. YJ confirmed that the linear growth is indeed unaffected by
gas compressibility, which we now see also remains weak in the non-linear
regime.

The gas fluctuations are consistent with the small Mach numbers in
\Tab{t:flow}, which are below (but near) the scale set by the pressure
supported velocity, $\eta v_{\rm K}$, with $\eta v_{\rm K}/c_{\rm s} = 0.05$ in
our simulations. Curiously, the range in \emph{radial} Mach numbers is
remarkably narrow for all the 2.5-D simulations, from $8.7 \times 10^{-3}$ to
$1.2 \times 10^{-2}$ (with the weakly turbulent, two-fluid run AA excluded).

\subsection{Particle Density Distribution}

\begin{figure}
  \includegraphics{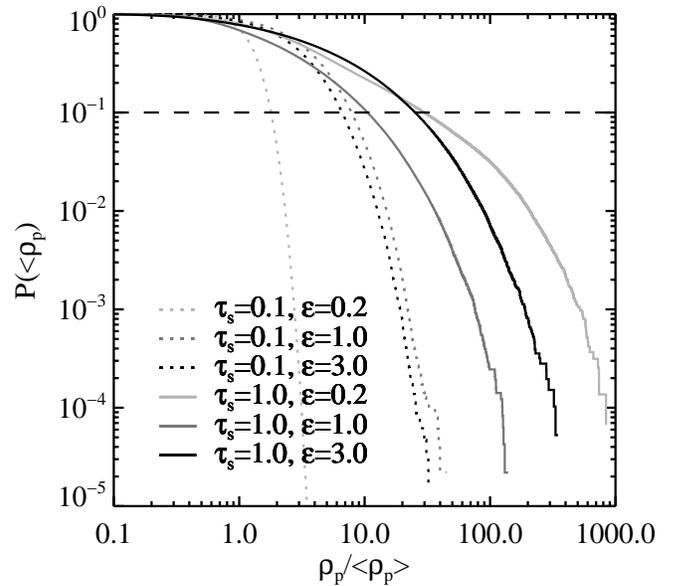}
  \caption{Cumulative particle density distributions. The curves show the
    fraction of particles with an ambient density $\geq \rho_{\rm p}$ (the
    dashed line indicates a 10\% border between typical and exceptional).
    Except for run AA ($\tau_s = 0.1, \epsilon = 0.2$) the majority of
    particles reside in clumps overdense by a factor of 2--10. A small fraction
    of particles experience extreme overdensities of nearly 1000.}
  \label{f:particlerhod}
\end{figure}
To get a clear picture of both typical and maximum particle overdensities,
\Fig{f:particlerhod} plots the cumulative distributions of particle density
during the saturated phase of the simulations. The distributions measure the
fraction of particles with ambient densities above a given value, and are
averaged over many snapshots to ensure adequate sampling. Particle densities
relative to the gas are readily obtained by multiplying the $x$-axis values by
$\epsilon = \langle\rho_{\rm p}\rangle/\langle\rho_{\rm g}\rangle$. Run BA
($\tau_s = 1.0$, $\epsilon = 0.2$) has the largest particle overdensities, of
nearly 1000, meaning $\rho_{\rm p}$ reaches nearly 200 times the gas density.
However since run BC ($\tau_s = 1.0$, $\epsilon = 3.0$) starts with a particle
density 15 times larger, it experiences larger peak values of $\rho_{\rm
p}/\langle\rho_{\rm g}\rangle \approx 900$. Curiously run BB ($\tau_s = 1.0$,
$\epsilon = 1.0$) is not an intermediate case but has smaller overdensities
relative to both particles and gas.

Particle concentration is more modest during the tightly coupled runs. Case AB
and AC have very similar particle overdensities, with an average $\delta_{\rm
p} \approx$ 2--3 and a peak $\delta_{\rm p} \approx$ 30. For case AA ($\tau_s =
0.1$, $\epsilon = 0.2$) the overdensities are negligible. It is remarkable (if
a bit mysterious) that the $\epsilon = 0.2$ runs give both the strongest (BA)
and weakest (AA) particle overdensities, depending on stopping time! 

\begin{figure*}
  \includegraphics{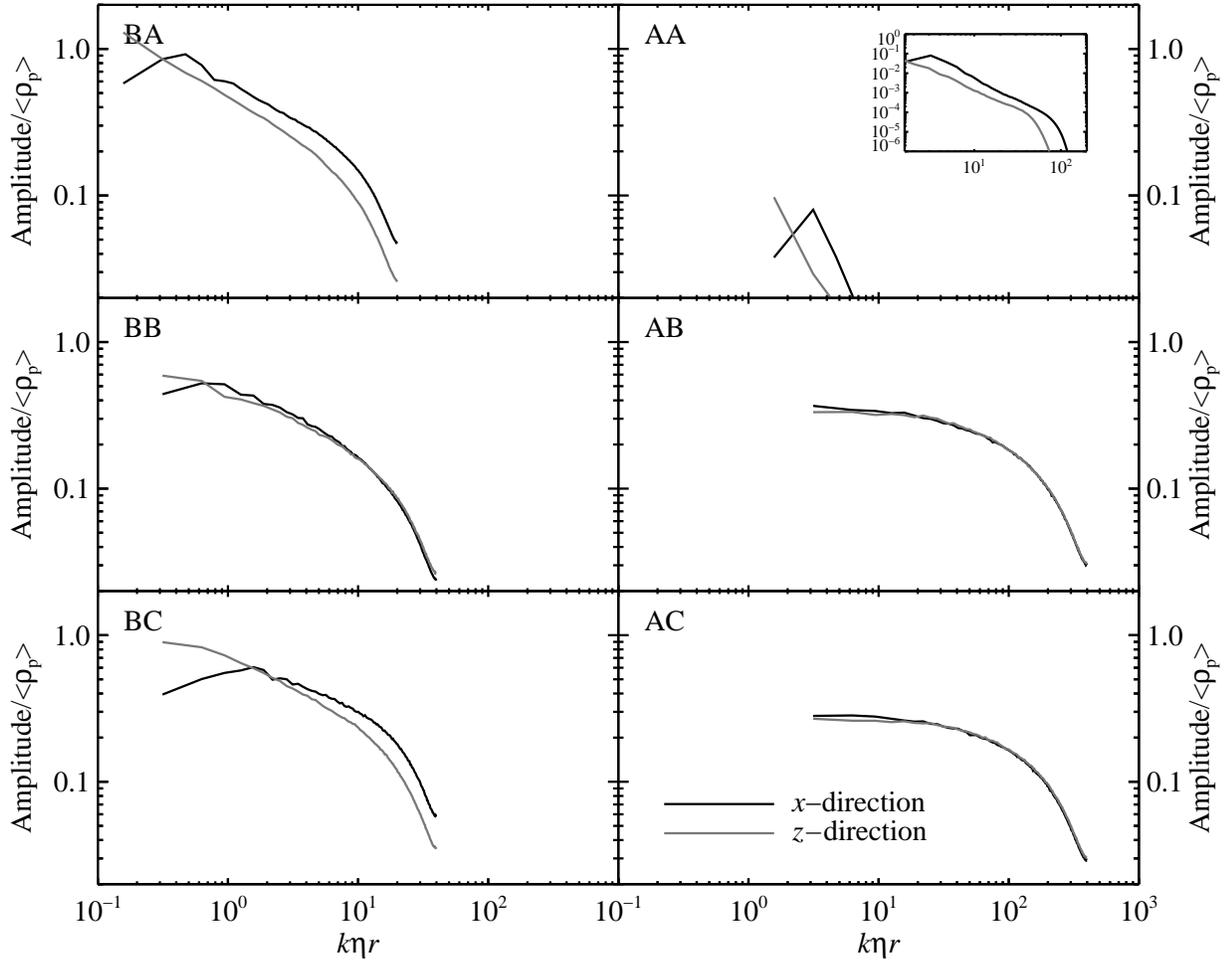}
  \caption{Power spectra of the bulk particle density, along the $x$-direction
    (black line) and the $z$-direction (gray line). The Fourier amplitudes are
    shown normalized with the mean density of particles in each simulation.
    Runs BA, BB and BC show clear peaks at large scales in agreement with the
    scale of the clumps seen in Figs.\ \ref{f:epsd_BA} and \ref{f:epsd_BB+BC},
    whereas the power in the tightly coupled runs AB and AC is largely
    isotropic and monotonically decreasing with decreasing wave length. Run AA
    is extremely top heavy with power almost exclusively at the few largest
    scales of the box (see insert).}
  \label{f:power_rhop}
\end{figure*}
Fourier spectra of the particle density are shown in \Fig{f:power_rhop}. The
absolute value of the Fourier amplitudes, normalized by the mean bulk density
of particles, has been averaged over many snapshots during the saturated
turbulent state of the simulations. Runs BA, BB and BC show clear peaks at
large scales in agreement with the scale of the clumps seen in Figs.\
\ref{f:epsd_BA} and \ref{f:epsd_BB+BC}, whereas the power in the tightly
coupled runs AB and AC is largely isotropic and monotonically decreasing with
decreasing wave length. Run AA is extremely dominated by the very largest
scales of the box.
\\ \\

\subsection{Correlation Times}\label{c:tcorr}

The residence time of particles in dense clumps affects the cosmogonical
processes, e.g.\ gravitational collapse or chondrule formation, that might
occur therein. For this purpose, we measure the time correlation function of
the ambient density, $\rho_{\rm p}^{(i)}$, experienced by particle $i$, 
\begin{equation}\label{eq:corr_time}
  C_\rho(t) = \langle\rho_p^{(i)}(t')\rho_p^{(i)}(t'+t)\rangle -
  \langle\rho_{\rm p}^{(i)} \rangle^2 \, ,
\end{equation}
from snapshots of the particle positions taken every $\Delta t=\varOmega^{-1}$
apart. The brackets indicate an average over time\footnote{Since averaging is
restricted to intervals $t$ apart, the largest $t$ considered is never more
than half the (non-linearly saturated) duration of the simulation.} and the
particles tracked (10\% of the total was more than sufficient for
convergence).  Subtraction of the mean squared $\rho_{\rm p}^{(i)}$ ensures
that positive (negative) values of $C_\rho(t)$ correspond to correlation
(anticorrelation), respectively.

\begin{figure}
  \includegraphics[width=8.7cm]{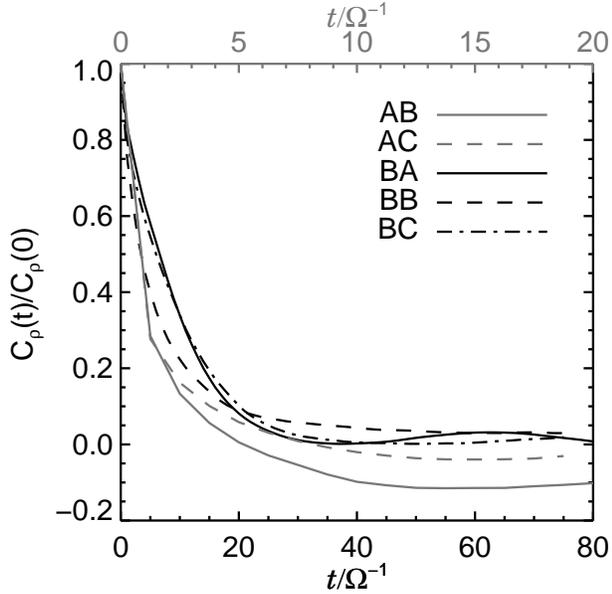}
  \caption{Time correlation functions for particle density indicate how long
    particles reside in dense clumps. Runs BA and BC have long correlation
    times of $\sim 7 \varOmega^{-1}$, while BB has a slightly shorter value of
    $3.5 \varOmega^{-1}$. These $\tau_s = 1.0$ runs use black lines and the
    lower black time axis. The tightly coupled runs, AB and AC (gray lines and
    the upper gray compressed time axis), have very short correlation times $<
    1 \varOmega^{-1}$.}
  \label{f:tcorr}
\end{figure}
\Fig{f:tcorr} plots the time correlation function for the saturated state of
the 2.5-D simulations. A characteristic correlation time, $t_{\rm corr}$, is
obtained when $C_{\rho}$ drops to half its peak value.\footnote{We considered
defining correlation functions and times only for particles initially residing
in overdense regions, but \Eq{eq:corr_time} is a quadratic measure that already
favors such regions. The simpler, more standard definition is sufficient for
our purposes.} Runs BA and BC have the longest $t_{\rm corr} \approx (6$--$7)
\varOmega^{-1}$. Run BA is the best sampled and shows a secondary peak past $t
= 60 \varOmega^{-1}$ indicating either periodicity or (more likely) secular
changes from the ongoing cascade and small clump numbers. Run BB enjoys a
shorter $t_{\rm corr} \approx 3.5 \varOmega^{-1}$, but $C_{\rho}$ does not
quite drop to zero, an indication that a fraction of particles remain in dense
regions.

The runs with tighter coupling of $\tau_s = 0.1$ (AB and AC) had quite short
$t_{\rm corr} \approx 0.7 \varOmega^{-1}$ (note the compressed time axis in
\Fig{f:tcorr} for these runs). The short correlation times are consistent with
the less pronounced clumping and lack of upward cascade when compared to the
marginally coupled runs. For run AC, $C_\rho$ is significantly negative for $t
> 20 \varOmega^{-1}$, indicating that particles avoid dense clumps after
leaving them.

It is clear from movies of the simulations that many clumps persist longer than
$t_{\rm corr}$, particularly in the marginally coupled $\tau_s = 1.0$ runs. The
particles that make up a clump continuously leak out downstream to the radial
drift flow and are replaced with new particles drifting in from upstream. The
mismatch between clump lifetime and density correlation time is evidence that
the clumps are a dynamical, collective phenomenon in the solid component,
rather than a persisting density enhancement. That situation might change with
the inclusion of the self-gravity of the solid particles, as this could cause
the clumps as a whole to collapse under their own weight, fragmenting perhaps
into gravitationally bound objects. We plan to include the self-gravity of the
particles in a future research project.

\subsection{Energetics of Clumping}\label{c:Eclump}

\begin{figure}
  \includegraphics[width=8.7cm]{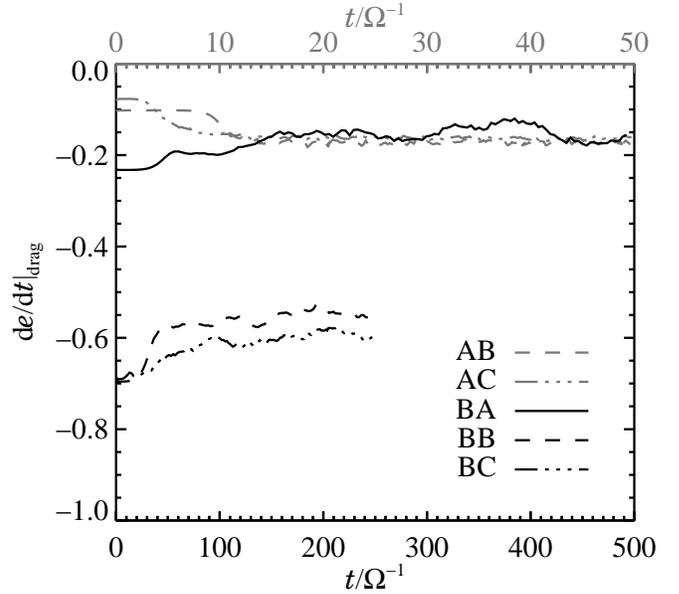}
  \caption{The energy dissipation rate [normalized to $\langle\rho_{\rm
    g}\rangle (\eta v_{\rm K})^2 \varOmega$] from drag forces between solids
    and gas. Marginally coupled runs BA, BB and BC (black curves) reduce the
    dissipation rate in the turbulent state, by shielding particles in dense,
    long-lived particle clumps. Tightly coupled runs AB and AC (gray curves,
    which follow the top gray time axis) show increased dissipation, since
    particle clumps are too short-lived to allow such shielding.}
  \label{f:dedragp_t}
\end{figure}

The growth of particle clumps shields solids from the full brunt of drag
forces, akin to the drafting practiced in bicycle pelotons. In YJ \S5.1 we show
that the rate of energy dissipation by drag forces,
\begin{equation} 
  \dot{\mathcal{E}}_\mathrm{drag} = -\rho_{\rm p} |\vc{v}_{\rm g}-\vc{v}_{\rm
p}|^2/\ts \, ,
\end{equation}
is diminished (brought closer to zero) by particle clumping in the laminar
state (at least for tight or marginal coupling). To determine the relevance of
this process for the saturated turbulent state, \Fig{f:dedragp_t} plots the
time evolution of the energy dissipation rate for the marginally coupled runs
(black time axis) and two of the tightly coupled runs (gray time axis). For the
tightly coupled runs (AB and AC) the dissipation actually becomes stronger
(more negative) in the saturated state, a consequence of increased relative
velocities in the turbulent state. Since these runs also have significant
overdensities, the lowering of $|\dot{\mathcal{E}}_\mathrm{drag}|$ is
apparently not a necessary condition for clumping. The short clump lifetimes
(see \Fig{f:tcorr}) is consistent with the inability to reduce dissipation by
drafting for $\tau_{\rm s} = 0.1$.

By contrast, all marginally coupled runs (BA, BB and BC) show diminished
dissipation in the non-linear state, more consistent with the analytic
expectations from clumps in a laminar flow. The longer correlation times and
the upward cascade into large clumps exhibiting bulk motion (particularly in
AB, see \S\ref{s:boulders}) foster the reduction of
$|\dot{\mathcal{E}}_\mathrm{drag}|$. The resulting particle overdensities are
significantly larger for these $\tau_{\rm s} = 1.0$ runs (see
\Fig{f:particlerhod}). Thus diminishing drag dissipation is not required to
generate particle overdensities, but this drafting mechanism can augment the
growth of dense clumps.

\section{Transport}\label{c:transport}

In this section we quantify the effect of the streaming turbulence on the
radial drift of particles, radial momentum transport and on the diffusive
mixing of solids.

\subsection{Radial Drift}\label{s:radialdrift}

We initially expected that streaming turbulence would reduce the radial
migration of particles, due to the pronounced particle clumping. The laminar
drift of particles slows as [see YJ equation (7c)]
\begin{eqnarray} 
 w_x^{\rm (NSH)} &=& -\frac{2 \tau_{\rm s}}
      {(1+\rho_{\rm p}/\rho_{\rm g})^2+\tau_{\rm s}^2} \eta v_{\rm K}
      \label{eq:NSHwx}  \\
&\rightarrow & - \left(\frac{\rho_{\rm p}}{\rho_{\rm g}}\right)^2 2 \eta v_{\rm K} \tau_{\rm s} 
\,\,\, \mathrm{for} \,\,\, {\rho_{\rm p}\over\rho_{\rm g}} \gg 1, \tau_{\rm s}
\end{eqnarray}
with increasing particle inertia. The results of the simulations are more
complicated since turbulent velocity fluctuations produce drift speeds that
deviate from local equilibrium.

\begin{figure}
  \includegraphics{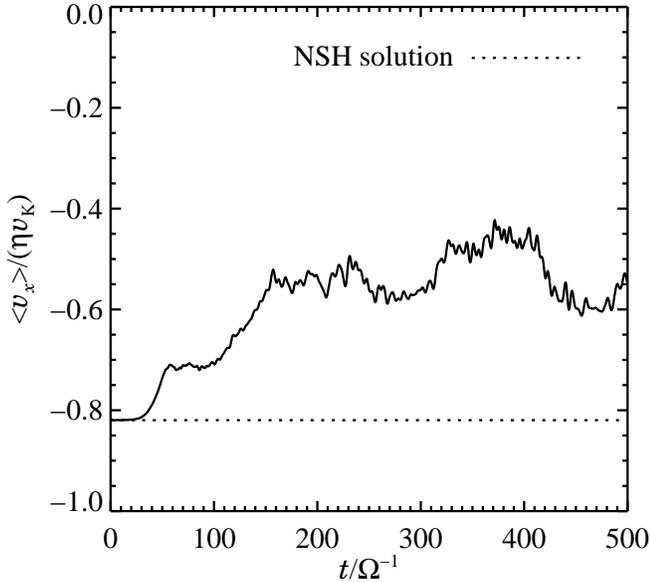}
  \caption{Evolution of the radial drift speed of solids during run BA,
    averaged over all 1,600,000 particles. Streaming turbulence slows the
    influx of solids by $40\%$ below the laminar drift speed (dotted line) on
    average, with significant temporal fluctuations that correlate with peaks
    in the maximum bulk density of particles (\Fig{f:rhopmax_t_t1.0}).}
  \label{f:vpxm_t}
\end{figure}
\Tab{t:flow} lists average radial drift velocities in the turbulent state,
along with the laminar equilibrium values from NSH. Radial drift decreases by
about $40 \%$ during run BA, as is also shown in \Fig{f:vpxm_t}. For the other
$\tau_{\rm s} = 1.0$ runs, BB displays a modest 15\% reduction while BC is
unchanged despite significant overdensities. The tightly coupled runs show
marked increases in drift speeds of 200\% for AB and 300\% for AC. Note that BA
has the fastest laminar drift (due to marginal coupling and low solids-to-gas
ratio), while AC (followed by AB) have the slowest laminar drift (because of
tight coupling and large particle inertia). While the same ordering of drift
speeds holds in the turbulent state, the range of speeds for different
parameter choices shrinks (i.e.\ the fastest slow down and the slowest speed
up). We examine this trend in detail below.

\begin{figure*}
  \vspace*{-4mm}
  \begin{center}
    \includegraphics[width=7.0cm]{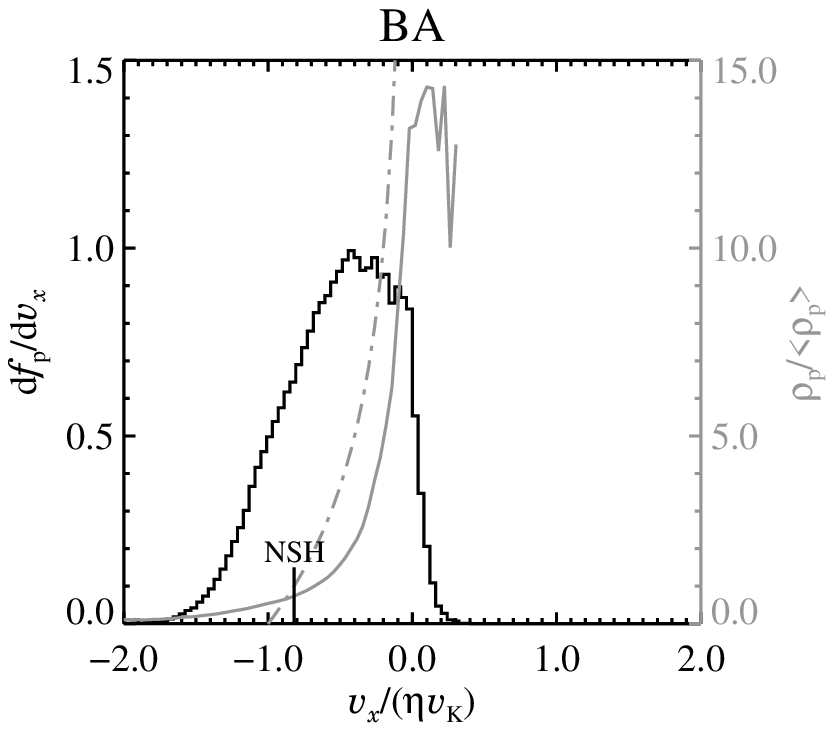}
    \includegraphics[width=7.0cm]{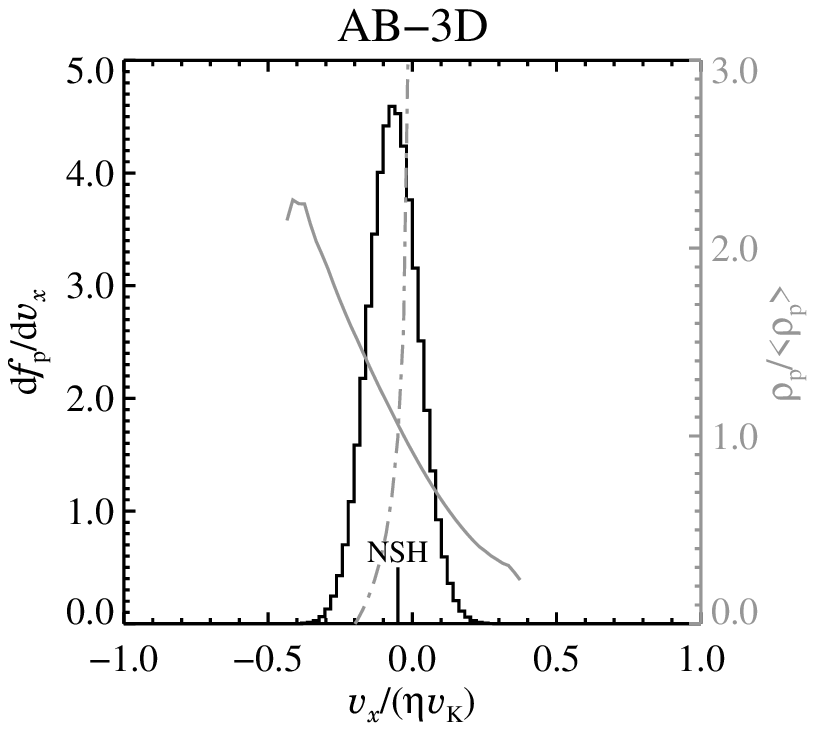}
    \includegraphics[width=7.0cm]{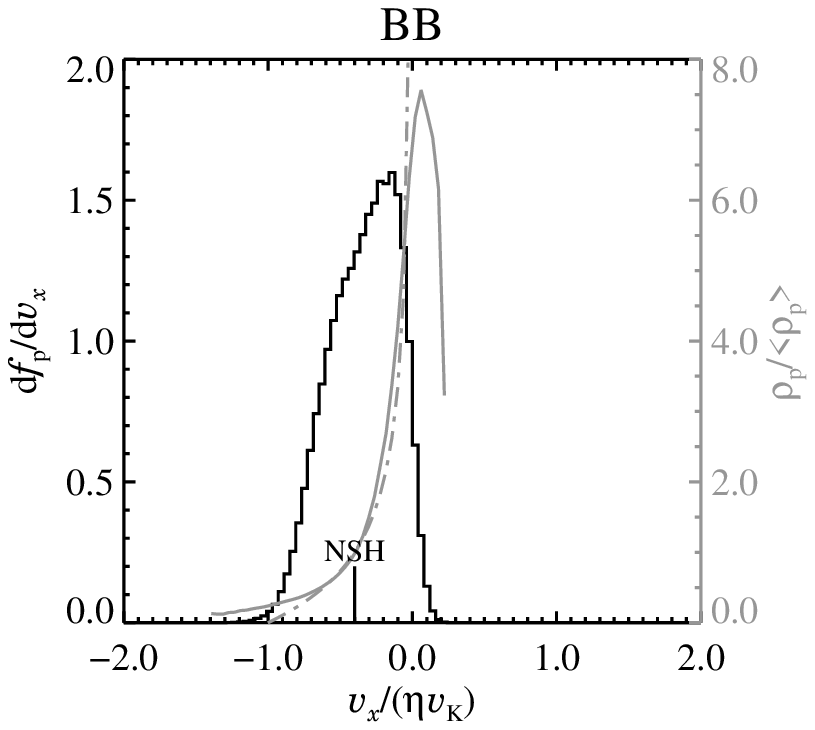}
    \includegraphics[width=7.0cm]{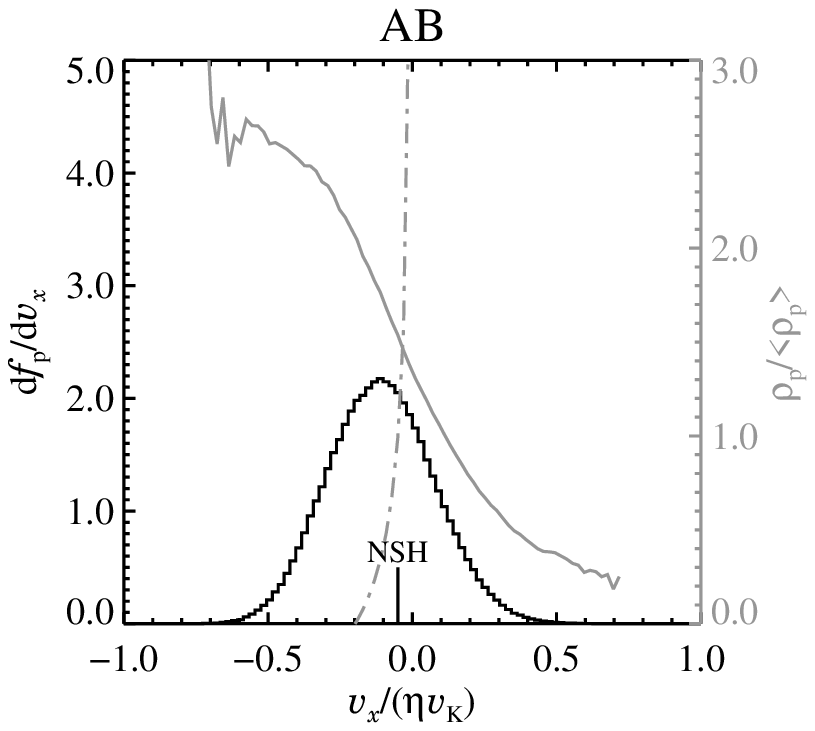}
    \includegraphics[width=7.0cm]{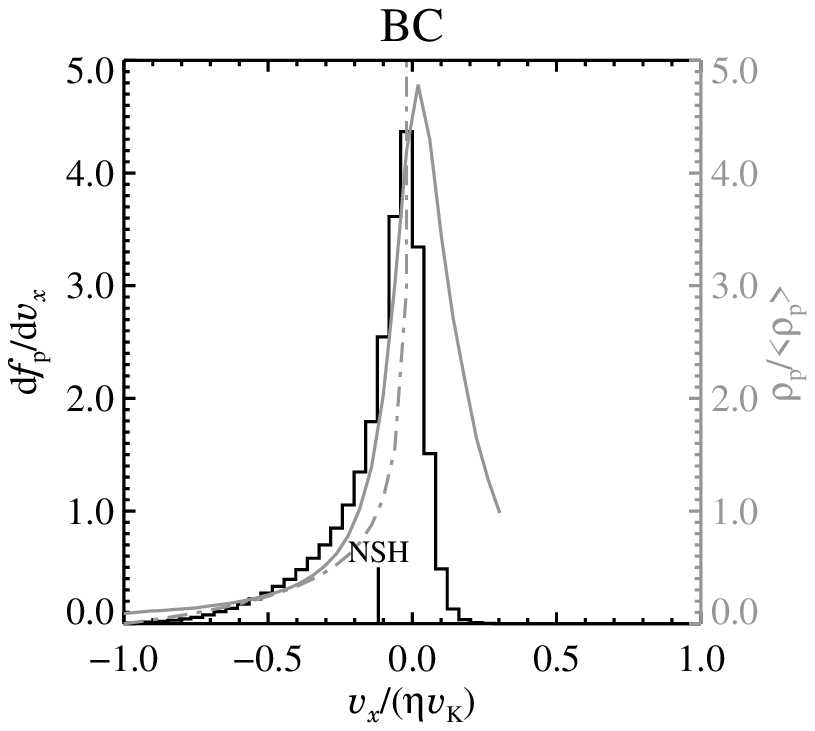}
    \includegraphics[width=7.0cm]{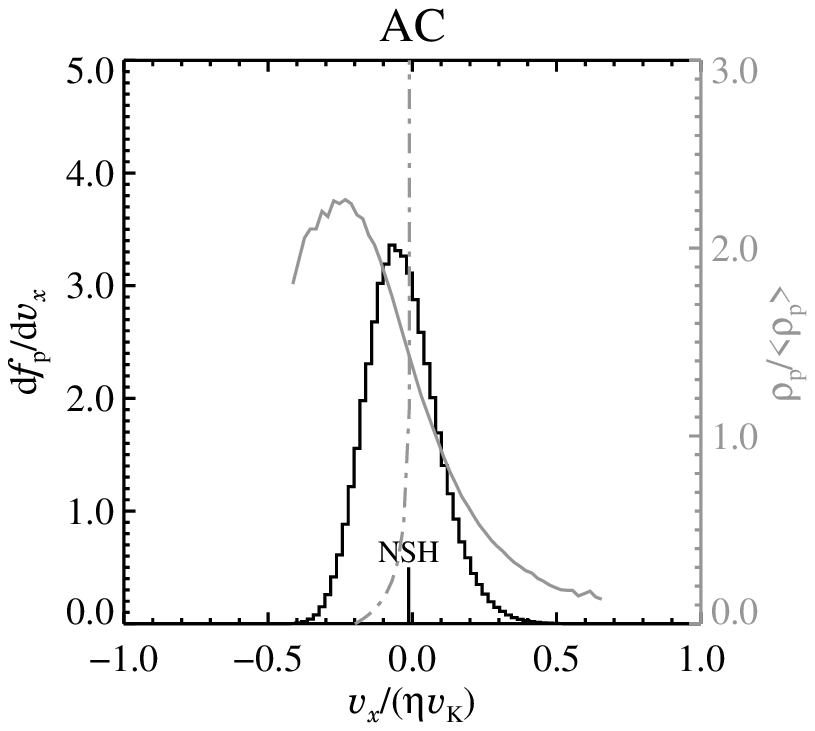}
  \end{center}
  \caption{Histograms of the fraction of particles with a given radial drift
    velocity $v_x$ in the turbulent state (black curves). The short vertical
    lines (labeled NSH) indicate analytical drift velocities in the initial
    state with no turbulence or clumping.  Marginally coupled (B*) runs show a
    slowing of the net drift speed, whereas tightly coupled runs (A*) produce
    faster infall. The gray lines (following the right y-axes) show the average
    particle density in each velocity bin. For reference the gray dash-dotted
    lines plot the laminar particle density vs.\ drift velocity relation. The
    B* runs display the expected decrease in drift speeds with increasing
    density, whereas the A* runs (surprisingly) follow the opposite trend.}
  \label{f:vpx_histogram}
\end{figure*}
\Fig{f:vpx_histogram} shows (with black histograms) the distribution of drift
velocities, averaged over time in the turbulent state, for six different runs.
For comparison, the location of the equilibrium drift velocity is indicated
with a short vertical line (labeled NSH). The full gray lines plot the average
ambient particle density for particles in a given velocity bin, and should be
compared to the dash-dotted gray lines that show the laminar relation between
particle density and drift velocity (from the inversion of eq.\
[\ref{eq:NSHwx}]). The laminar drift velocities have a finite range from 0 for
infinite particle densities to the single-particle case, $w_x^{\rm (min)} =
-\eta v_{\rm K}$ for $\tau_{\rm s} = 1.0$ and $w_x^{\rm (min)} \simeq -0.2 \eta
v_{\rm K}$ for $\tau_{\rm s} = 0.1$. The actual velocity range extends beyond
these limits in the turbulent state. 

First consider the marginally coupled runs (left column of
\Fig{f:vpx_histogram}). The velocity distribution is non-Gaussian with a clear
negative skewness (velocities drop sharply at the right side of the Gaussian,
with a more gradual decline toward negative velocities). The gray lines show
the expected trend of slower inward drift for higher ambient densities. For
particles moving radially outward with $v_x>0$, the average particle density
drops with increasing speed. This is reasonable behavior since low density
particle clumps can more readily be fed angular momentum and pushed outward by
gas fluctuations.\footnote{In the absence of fluctuations and with an outwardly
decreasing pressure, particles only drift inwards.} The extended tails of
fast-drifting material at low densities are responsible for the modest (or
non-existent for BC) reduction of drift velocities, despite the slowing, or
even reversal, of motion in overdense regions.

Now consider the tightly coupled runs in the right column of
\Fig{f:vpx_histogram}. The velocity distributions are nearly Gaussian and
extend well beyond the range of laminar drift velocities, indicating that
turbulent fluctuations dominate. The peaks are shifted leftward, which produces
the higher turbulent drift speeds of \Tab{t:flow}. The gray curves plot the
astounding fact that overdense regions drift in faster, a reversal of the
laminar trend. This is seen in the movie of AB where dense clumps snake their
way inwards while underdense diffuse material races out (the snake patterns are
visible in \Fig{f:epsd_AB} as well). 

\subsubsection{Effective Drag on Clumps}

The tendency for faster migration of dense clumps for $\tau_{\rm s} = 0.1$ can
be understood as a consequence of an effective, macroscopic drag force acting
on the clumps. The gas inside the clump is tied to the clump, but exterior gas
passes freely around the surface, exerting an effective drag. If the effective
friction time of the clump is closer to unity than the original $\tau_{\rm s}$,
then the dense clump will behave more like a marginally coupled solid and drift
inward faster. This collective drag effect is similar to the plate drag model
of Ekman layers on the surface of particle subdisks \citep{gw73,gp00}. 

We estimate the friction time $\tau_{\rm f}^{\rm (eff)}$ of a clump of radius
$R_{\rm clump}$ as the time required to encounter its own mass, $M_{\rm
clump}$, in gas.\footnote{This is the valid criterion for high Reynolds number,
turbulent drag.} That gives for 3-D clumps (2-D clumps give the same final
scaling)
\begin{equation}
  \tau_{\rm f}^{\rm (eff)} \sim \frac{M_{\rm clump}}{\rho_{\rm g} R_{\rm clump}^2 \Delta v} \sim
\frac{\rho_{\rm p}}{\rho_{\rm g}} \frac{R_{\rm clump}}{\Delta v}\, ,
\end{equation}
where $\rho_{\rm p}$ is the bulk particle density inside the clump and $\Delta
v$ is the speed of the clump relative to the gas. Multiplying each side by
$\varOmega$ yields
\begin{equation}
  \varOmega \tau_{\rm f}^{\rm (eff)} \sim
      \frac{\rho_{\rm p}}{\rho_{\rm g}} \frac{R_{\rm clump}}{\eta r}
      \frac{\eta v_{\rm K}}{\Delta v} \, .
\end{equation}
If we dare test this heuristic hypothesis, reading the size of the clumpy
plateaus from Figs.\ \ref{f:epsd_AB} and \ref{f:epsd_AA+AC} and the bulk
density and speed of the dense clumps relative to the gas from
\Fig{f:vpx_histogram}, we find for run AB the values $\rho_{\rm p}/\langle
\rho_{\rm g} \rangle \approx 2.5$, $R_{\rm clump}/(\eta r) \approx 0.1$,
$\Delta v \approx \eta v_{\rm K}$ (the velocity difference between high density
material and low density material in \Fig{f:vpx_histogram}), corresponding to
an effective friction time of $\varOmega \tau_{\rm f}^{\rm (eff)} \approx 0.25$
Run AC has $\rho_{\rm p}/\langle \rho_{\rm g}\rangle \approx 6$, $R_{\rm
clump}/(\eta r) \approx 0.05$, $\Delta v \approx \eta v_{\rm K}$, yielding a
very similar value of $\tau_{\rm f}^{\rm (eff)} \approx 0.3$. Thus our crude
estimates show that the clumps couple aerodynamically to the gas more loosely
than the individual particles do, explaining at least qualitatively the faster
drift of dense clumps and the increase in drag dissipation, two surprising
features of the $\tau_{\rm s} = 0.1$ runs.

\subsection{Momentum Flux}

The radial flux of orbital momentum, $\mathcal{F}_{\mathcal{L},x}=\rho_{\rm g}
u_x u_y + \rho_{\rm p} w_x w_y$, and its contribution to disk heating are
discussed in YJ \S 5. For laminar flow the drag equilibrium between solids and
gas gives (equation 18b of YJ) 
\begin{equation}
  \mathcal{F}_{\mathcal{L},x}^{\rm (NSH)}= -2 \tau_{\rm s}^3 \rho_{\rm p}
      \left[ \frac{\eta v_{\rm K}}{(1+\epsilon)^2+\tau_{\rm s}^2} \right]^2\, .
\end{equation}
The \emph{inward} transport of angular momentum follows from the the slower
rotation of the outgoing gas and the faster rotation of the incoming
particles.  The values for $\mathcal{F}_{\mathcal{L},x}$ in the saturated
turbulent flow are given in \Tab{t:turbulence} and are decomposed as
$\mathcal{F}_{\mathcal{L},x}= \mathcal{F}_{\mathcal{L},x}^{\rm
(NSH)}+\mathcal{F}_{\mathcal{L},x}^{\rm (turb)}$, i.e.\ the laminar value and
changes caused by turbulence. If the turbulence were driven by orbital shear,
which releases free energy via outward angular momentum transport,
$\mathcal{F}_{\mathcal{L},x}^{\rm (turb)}$ would be positive. Instead, most
runs have $\mathcal{F}_{\mathcal{L},x}^{\rm (turb)} < 0$, which is physically
allowed since work done by the global pressure gradient powers streaming
turbulence. Only run BA (and BA-3D) has $\mathcal{F}_{\mathcal{L},x}^{\rm
(turb)} > 0$, but the net angular momentum flux is still inward. Thus in all
our simulations, angular momentum transport acts to take kinetic energy out of
the motion, at the rate $\dot{\mathcal{E}}_{\mathcal{L}} = (3/2)\varOmega
\mathcal{F}_{\mathcal{L},x} < 0$ (see YJ \S5). As in any shearing box
simulation with (shear) periodic boundary conditions, momentum fluxes are
divergence-free constants, which prohibits secular evolution. Global
simulations are needed to fully investigate the role of ``backwards'' angular
momentum transport from streaming turbulence on disk evolution.

\subsection{Turbulent Diffusion}\label{s:turbdiff}

The turbulent mixing of particles is usually modeled as a diffusive process in
which particle motions are described by a random walk for large length-scales
and over long time-scales. We provide here best fits to the diffusion
coefficients in the radial and vertical directions. Since the motion of
particles is very complex, and furthermore the particles are not passive
contaminants but the cause of turbulence, we also test the validity of the
diffusion approximation.

\begin{figure}
  \includegraphics[width=8.7cm]{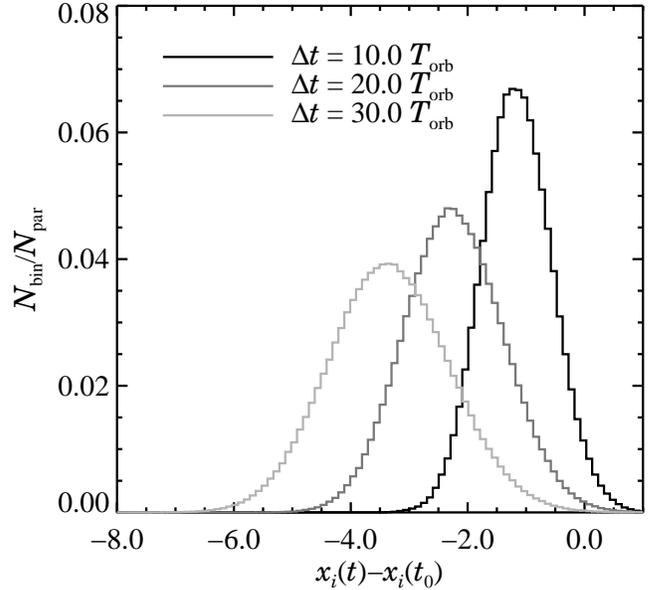}
  \caption{Spot the platypus. The histograms plot the radial distance that
    particles in run BA have traveled since the reference time of $t_0=20
    T_{\rm orb}$. The curve moves inward due to the radial drift, while
    spreading as a random walk with a Gaussian width $\sigma$ that increases as
    the square root of time. The Gaussians are slightly platykurtic, or
    flat-tailed, due to a population of solid particles that experience
    decreased diffusion in the massive particle clumps seen in \Fig{f:epsd_BA}.}
  \label{f:sigma2_t}
\end{figure}

We track the deviation of particle positions, $x_i(t)$ and $z_i(t)$, from their
positions at an initial time $t_0$ when turbulence has already developed. Here
we are not concerned with the net radial drift of particles, but the spreading
of the distribution $x_i(t)-x_i(t_0)$ (and similarly for vertical motions,
although no systematic motion over long time-scales is expected in this
direction). Particles are allowed to move greater distances than the box size
by deconvoluting any particles that were transferred over the periodic
boundaries by the code. For pure diffusion, the distribution tends to a
Gaussian with a variance, $\sigma^2$, that grows linearly with
time.\footnote{Since particles are allowed to cross the periodic boundaries, at
late times different portions of the distribution will overlap. This just means
that only turbulent scales up to a certain length scale are considered,
something that should not significantly affect the integrity of the
measurements.} An example of how the particles spread out with time is shown in
\Fig{f:sigma2_t} for radial mixing in the BA simulation. The diffusion
coefficients, $D_x$ and $D_z$, are extracted as
\begin{equation}
  D_{x,z} \equiv {1 \over 2}{\partial \sigma^2_{x,z} \over \partial t} \, .
\end{equation}
The best fit diffusion coefficients are listed in \Tab{t:turbulence}, using the
standard normalization by $c_{\rm s}^2 \varOmega^{-1}$. The dimensionless
diffusion coefficients lie in the interval $10^{-7}\ldots10^{-2}$, ranging from
extremely small up to values that are comparable to the diffusion caused by
magnetorotational turbulence \citep{Carballido+etal2005,JohansenKlahr2005}. For
the smaller $\tau_s = 0.1$ particles the diffusion is quite weak, $< 5 \times
10^{-5}$. This is because more tightly coupled particles trigger weaker
turbulence with smaller length scales, a result consistent with smaller linear
growth rates and wavelengths for lower $\tau_s$ (see YG). Run BA ($\tau_s =
1.0, \epsilon = 0.2$) exhibits anomalously large diffusion, especially in the
vertical direction, $D_z \approx 0.01$. This is due to the significant bulk
motion of elongated clumps (discussed in \S\ref{s:boulders}).

The upper limit for the diffusion appears to be set by the characteristic
length and velocity scales, $\eta r$ and $\eta v_{\rm K}$, to be $D \lesssim
\eta^2 v_{\rm K} r \approx \eta c_{\rm s}^2/\varOmega$, i.e.\ $D \lesssim \eta
= 5 \times 10^{-3}$ when non-dimensionalized. Indeed even the extreme $D_z$ in
run BA only violates this order of magnitude criterion by a factor of three.
As a consistency check on the diffusion coefficients, $D_{x,z} \approx \delta
w_{x,z}^2t_{\rm corr}$ is obeyed within a factor of a few, for random
velocities, $\delta\vc{w}$, from \Tab{t:flow} (for the gas, but particle values
are not more than $\sim$ 10\% different) and the correlation times, $t_{\rm
corr}$, from \Fig{f:tcorr}.

It will be interesting to compare these results to stratified disk models with
self-consistent vertical settling, where the relevant parameters are $\tau_s$
and the solids-to-gas \emph{surface} density ratio (instead of $\epsilon$).

\subsubsection{Validity of the Diffusion Approximation}

We performed several tests of the diffusion approximation. The time variation
of the diffusion coefficients should be small, and especially should lack an
overall deviation from $\partial \sigma^2/\partial t \propto
\mathrm{constant}$. This was true for most runs, as indicated by the error bars
on the diffusion coefficients in \Tab{t:turbulence}. The two exceptions were
again run BA, which exhibited large fluctuations due to the interactions
between a few large particle clumps, and the two-fluid run AA. This run was
seeded with tracer particles, following the velocity field of the solid fluid,
in order to be able to use the random walk approach to measure diffusion. The
tracer particles exhibited extremely small diffusion with a huge fluctuation
interval, an effect of the weak non-linear state of run AA where periodic bulk
motion of a few clumps dominates over random motion (see \Fig{f:epsd_AA+AC}).
Particles spread out and gather again in a way that is distinctly not a random
walk, but over longer time-scales the particle distribution still spread out as
a Gaussian. The enormous error interval indicates that the turbulent transport
is not like diffusion on short time-scales.

We also tested Gaussianity by measuring the skewness and kurtosis of the
particle displacement distributions. Most runs were fairly Gaussian, except for
a modest skewness, $\sim 10\%$, in the radial (and not vertical) distributions,
which is readily explained by the interaction of the radial drift flow with
clumps. The BA run exhibited a kurtosis of $-0.5$, i.e.\ slightly platykurtic
or small-tailed (see \Fig{f:sigma2_t}), consistent with transport influenced by
bulk motions, and not just a random walk. Modeling turbulent transport as
diffusion is under all circumstances only an approximation. Still, the
turbulent diffusion coefficient is a good measure of the time-scale on which
solid particles are mixed by the turbulent motion.

\section{Summary}\label{c:conc}

In this paper we have shown that solid particles can trigger turbulence in
gaseous protoplanetary disks via the streaming instability and thereby cause
their own clumping. We have ignored a number of complicating effects. Most
critical is perhaps the lack of vertical gravity, but we believe it was
instructive to see how the streaming instability evolves in a pure model that
has exact linear solutions first. We plan to include both vertical gravity and
the self-gravity of the solids in a future research project. A distribution of
particle sizes and physical collisions between particles have also been
ignored, even though coagulation, fragmentation, and collisional cooling are
likely relevant in dense particle clumps. As the complex behavior of our simple
model system shows, significant progress on basic physical processes can be
made before the ``kitchen sink" approach is required.

The most striking consequence of streaming turbulence is the growth of
overdense particle clumps without self-gravity. This effect was previously seen
in the non-linear simulations of particle settling and Kelvin-Helmholtz
turbulence by \cite{jhk06}. In both that work and this one, clumping can be a
self-propagating phenomenon. The increased inertia in dense clumps decreases
their drift speeds, creating local ``traffic jams." We saw this behavior for
marginally coupled solids, which developed the largest relative overdensities,
above 100, with an upward cascade to long-lived, vertically elongated
filaments.  Marginally coupled solids have the highest radial drift speed and
are known to exhibit the most pronounced drag-related phenomena, so it is not
surprising that marginal coupling also gives rise to the most dramatic
streaming turbulence. While clumping may not in itself explain how to keep
large amounts of marginally coupled particles at large orbital distances
\citep{Wilner+etal2000}, it does provide a rigorous prediction that the spatial
distribution of such solids will not be smooth, but will vary on scales of
around one gas scale-height.

A qualitatively different clumping behavior was seen for smaller, more tightly
coupled solids. Overdensities were lower, in the tens, and clumps were smaller
scale and short-lived. To extend the analogy, these runs appeared more like a
game of bumper cars than a full-scale traffic jam. The biggest surprise was the
complete reversal of the laminar relation between drift speed and particle
density. Dense clumps actually fell in faster than particles in voids for the
tightly coupled solids. Our heuristic explanation is that robust clumps can
withstand turbulent boundary layer flows that sap their angular momentum, as in
an Ekman layer flow. A similar explanation has been applied to the surfaces of
particle sub-disks as the plate drag {\it ansatz} \citep{gw73, gp00,stu03}. The
run with the tightest coupling and lowest initial solids-to-gas density ratio,
and consequently the lowest linear growth rate, developed very meek non-linear
density fluctuations. Thus non-linear clumping appears to require either
marginal coupling or a moderately large background solids-to-gas ratio of
around unity or higher. Further studies of the streaming instability for
smaller particles, such as chondrules, would be interesting, but are
computationally costly (see \S\ref{s:tight}).

It is hardly surprising that the solids-to-gas density ratio strongly affects
the non-linear state since the streaming instability relies on particle
feedback to influence gas dynamics. However, the sharp transition across
particle-gas equality is remarkable, especially for tight coupling -- the weak
streaming instabilities in the gas-dominated regime become explosive once the
solids-to-gas ratio reaches unity. \citet{ys02} argued that this threshold also
sets a limit to the quantity of solids that can be stirred by the
Kelvin-Helmholz instability.  Vigorous turbulence (if stronger than particles
themselves can stir) could prevent the accumulations of such high midplane
particle densities. There is little doubt, however, that dramatic events occur
whenever particle densities reach that of the gas. 

Planetesimal formation models generally involve either high particle densities,
in a gravitational collapse scenario, or efficient coagulation to particle
sizes for which radial drift is no longer a problem.  Both scenarios involve
conditions -- high particle densities and/or marginal drag force coupling --
where streaming instabilities can abet further growth towards planetesimals by
generating overdense clumps in the particle component. 

\acknowledgments

The authors would like to thank Jim Stone, Jeremy Goodman, Hubert Klahr and
Patrick Glaschke for inspiring discussions and an anonymous referee for
comments and suggestions that helped improve the original manuscript. A.J.\ is
grateful to the Annette Kade Graduate Student Fellowship Program at the
American Museum of Natural History. A.Y.\ acknowledges support NASA grant
NAG5-11664 issued through the Office of Space Science. 

{\it Facilities:} \facility{RZG (PIA)}

\end{document}